\documentclass[lettersize,journal]{IEEEtran}
\usepackage{amsmath,amsfonts,amsthm}
\usepackage{algorithmic}
\usepackage{algorithm}
\usepackage{array}
\usepackage[caption=false,font=normalsize,labelfont=sf,textfont=sf]{subfig}
\usepackage{textcomp}
\usepackage{stfloats}
\usepackage{url}
\usepackage{verbatim}
\usepackage{graphicx}
\usepackage{cite}
\usepackage{booktabs}
\usepackage{multirow}
\usepackage{makecell}
\usepackage{color}
\usepackage{CJKutf8}
\usepackage{pifont}
\usepackage{colortbl}
\newcolumntype{H}{>{\setbox0=\hbox\bgroup}c<{\egroup}@{}}
\usepackage{bm}
\usepackage{bbm}
\usepackage{pythonhighlight}
\usepackage{mathrsfs}
\usepackage{mathrsfs}
\usepackage{MnSymbol}
\usepackage{dashrule}
\usepackage{tikz}
\usepackage{mathtools}
\usepackage{amsthm}

\usepackage{adjustbox}
\usepackage{pifont}

\usepackage[utf8]{inputenc} 
\usepackage[T1]{fontenc}    
\usepackage{hyperref}       
\usepackage{url}            
\usepackage{booktabs}       
\usepackage[table]{xcolor} 
\usepackage{bm} 
\usepackage{amsfonts}       
\usepackage{nicefrac}       
\usepackage{microtype}      
\usepackage{xcolor}         
\usepackage{siunitx} 
\usepackage{subcaption}
\usepackage{wrapfig}
\usepackage{multirow}
\usepackage{multicol}
\usepackage{booktabs}       
\usepackage{graphicx}
\usepackage{amsmath}
\usepackage{array}
\usepackage[table,xcdraw]{xcolor}
\usepackage{color}
\usepackage{algorithm}
\usepackage{algorithmic}
\usepackage{hyperref}     
\usepackage{cleveref}     
\usepackage[numbers]{natbib}  
\usepackage{cite}
\definecolor{maroon}{cmyk}{0,0.87,0.68,0.32}
\definecolor{bamboo}{cmyk}{0.4,0,0.3,0}
\definecolor{apple}{cmyk}{0.41,0.4,0.76,0}
\definecolor{jialingshui}{cmyk}{0.47,0,0.49,0}
\definecolor{sea}{cmyk}{1,0.67,0.16,0.03}
\newcommand{\Romannum}[1]{\uppercase\expandafter{\romannumeral #1}}

\usepackage{booktabs}   

\usepackage{soul}          
\usepackage{xcolor}        
\newcommand{\best}[1]{\textbf{#1}}          
\newcommand{\second}[1]{\underline{#1}}     
\newcommand{\Updated}{\includegraphics[height=1.2em]{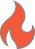}}
\newcommand{\Frozen}{\includegraphics[height=1.2em]{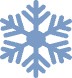}}

\definecolor{orange}{RGB}{247, 224, 213}

\begin{document}

\title{LangRetrieval: Language-Guided Self-Evolving Satellite-to-Radar Retrieval via CSI-Driven Reward}

\author{Chunlei~Shi, Junming~Hou, Yi-Lin~Wei, Jiong~Wang, Yecheng~Zhang, Yichao~Dong, \\Wenqi ~Ren,~\IEEEmembership{Senior Member,~IEEE}, and Dan~Niu,~\IEEEmembership{Member,~IEEE}

\thanks{This work was supported by the  Heavy Rainfall Research Foundation of China (No. BYKJ2025M14), China Meteorological Administration Xiong\textrm{'}an Atmospheric Boundary Layer Key Laboratory (No. 2025LABL-B12), and by the National Natural Science Foundation of China (62374031, 62331009), and by NSFC-Jiangsu Province (BK20240173). (\textit{Chunlei Shi, and Junming Hou have contributed equally to this work. Corresponding author: Wenqi Ren and Dan Niu.})}
\thanks{Chunlei Shi, Junming Hou, Yichao Dong and Dan Niu are with the Department of Automation, Southeast University, Nanjing 210096, China, and also with the State Key Laboratory of Millimeter Waves, School of
Information Science and Engineering, Southeast University, Nanjing 210096,
China (e-mail: 230238514@seu.edu.cn, junming\_hou@seu.edu.cn, danniu1@163.com).}
\thanks{Yi-Lin Wei is with the School of Computer Science and Engineering, Sun Yat-sen University, Guangzhou 510006, China. Jiong Wang is with the Department of Information Science and Technology, Fudan University, Shanghai 200433, China. Yecheng Zhang is with the Department of Architecture, Tsinghua University, Beijing 100084, China. Wenqi Ren is with the School of Cyber Science and Technology, Shenzhen campus of Sun Yat-sen University, Shenzhen 518107, China (e-mail: renwq3@mail.sysu.edu.cn).}}

\markboth{Journal of \LaTeX\ Class Files,~Vol.~14, No.~8, August~2021}%
{Shell \MakeLowercase{\textit{et al.}}: A Sample Article Using IEEEtran.cls for IEEE Journals}


\maketitle

\begin{abstract}
Satellite-to-radar (S2R) retrieval estimates ground radar precipitation from geostationary satellite observations, providing a critical solution for precipitation monitoring in radar-sparse regions. 
However, S2R retrieval is intrinsically ill-posed: similar cloud-top radiances can correspond to distinct precipitation regimes, storm organizations, and surface intensities, 
which are difficult to uniquely determine the underlying meteorological state from local spectral cues alone.
Meteorological semantics offer complementary scene-level information that can help resolve this ambiguity. 
Yet existing static semantic conditioning is often insufficient, as externally predefined semantics cannot adapt to dynamic convective scenes or align with retrieval objectives.
To this end, we propose LangRetrieval, a language-guided conditional flow matching (CFM) framework that establishes a closed-loop optimization mechanism between meteorological semantics and retrieval accuracy.
Specifically, LangRetrieval consists of two core components:
(i)~Semantic Warm-up: structured meteorological attributes are injected into the CFM backbone through cross-attention conditioning, enabling continuous semantic guidance throughout the generation trajectory; and
(ii)~Self-Evolving Semantic Optimization: a lightweight attribute policy is first initialized from vision-language model annotations and subsequently refined via Group Relative Policy Optimization (GRPO) using multi-threshold Critical Success Index (CSI) rewards, enabling semantic generation to evolve directly toward improved retrieval accuracy.  
Experiments on both the FY-4B and public SEVIR datasets demonstrate that LangRetrieval achieves state-of-the-art performance with a favorable performance-efficiency trade-off.
Beyond performance gains, our framework transforms meteorological semantics from static auxiliary input into a task-dependent optimizable variable, providing a promising solution for language-guided S2R retrieval.
\end{abstract}


\begin{figure}[ht]
    \centering
    \includegraphics[width=0.5\textwidth]{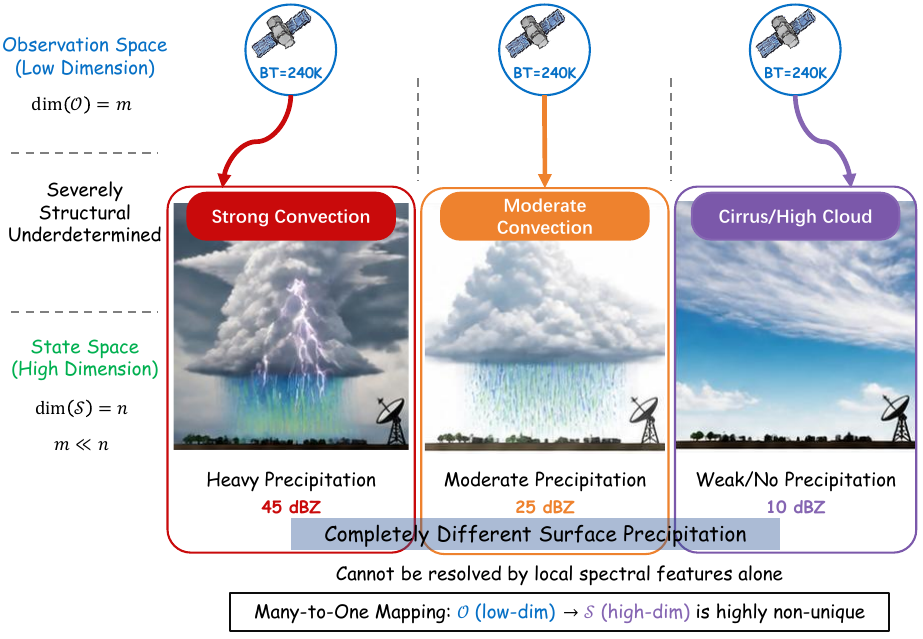}
    \vspace{-3mm}
    \caption{The many-to-one ambiguity in S2R retrieval: similar cloud-top satellite signatures, such as comparable brightness temperatures, can correspond to markedly different radar-observed precipitation intensities near the surface.}
    \label{fig:observation_state_gap}
\end{figure}

\section{Introduction}
Accurate prediction and monitoring of severe convective weather systems and extreme precipitation events are essential for public safety and disaster mitigation, with significant implications for fields such as weather forecasting, flood early-warning systems, and agricultural risk management~\cite{lin2025alphapre, gao2025lmcast, yu2025pimmnet, allen2025end,niu2025m4caster}.
Ground-based weather radar provides the high-resolution precipitation measurements for detecting and tracking these systems~\cite{xu2026synweather}, yet its coverage is largely restricted to populated land areas, leaving vast radar-sparse regions vulnerable to undetected extreme weather. 
In contrast, geostationary satellites such as Himawari-8, GOES-R, and FY-4B offer continuous global observation at 10--15~minute intervals with 1~km spatial resolution, making them ideal candidates for supplementing radar coverage gaps. 
Consequently, satellite-to-radar (S2R) retrieval, which estimates ground-based radar precipitation from satellite observations, emerges as a critical capability for extending convective detection to radar-sparse regions and providing timely warnings for extreme weather events~\cite{shi2026wavec2r}.
However, despite its practical importance, S2R retrieval is not a conventional image-to-image mapping problem.
Satellite sensors mainly observe cloud-top radiances, whereas radar reflects near-surface precipitation structures; this observation-state mismatch makes S2R intrinsically ill-posed, since near-identical cloud-top signatures may correspond to markedly different precipitation states (Fig.~\ref{fig:observation_state_gap}).

\begin{figure*}
    \centering
    \includegraphics[width=\textwidth]{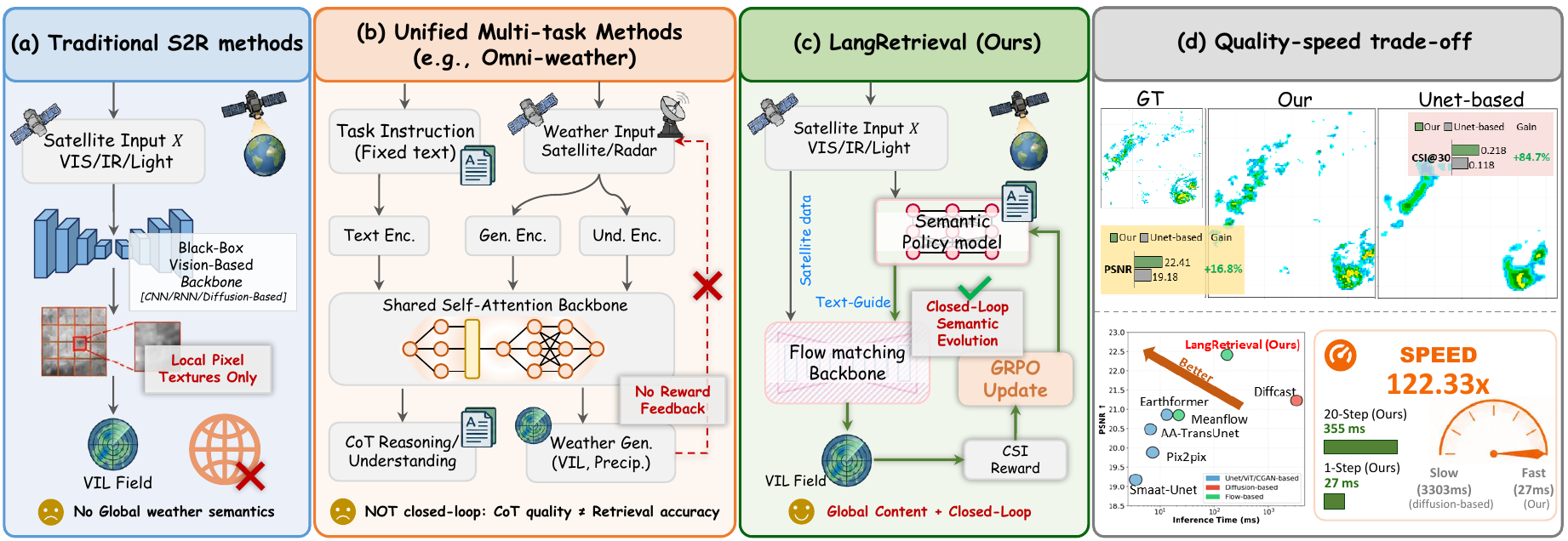}
    \vspace{-3mm}
    \caption{Comparison of S2R retrieval paradigms.
    (a)~\textit{Traditional methods} directly regress VIL from multi-spectral satellite inputs without meteorological semantic guidance, causing cirrus false alarms and extreme-precipitation underestimation.
    (b)~\textit{Unified multi-task methods} (e.g., Omni-Weather) introduce language via task instructions and supervised CoT reasoning, but the semantic signal remains open-loop and not optimized by retrieval metrics.
    (c)~\textit{LangRetrieval (Ours)} closes this loop: a learnable attribute policy steers the CFM backbone via attention-guided and is self-evolved by GRPO with multi-threshold CSI rewards.
    (d)~\textit{Quality-speed trade-off}, showing that LangRetrieval balances retrieval accuracy, perceptual quality, and inference efficiency.
    }
    \label{fig:motivation}
    \vspace{-4mm}
\end{figure*}

Over the past few years, the S2R community has witnessed rapid progress in deep learning methodologies, ranging from CNN-based architectures~\cite{trebing2021smaat,wu2025hvmunet} and transformer-based approaches~\cite{gao2022earthformer} to generative models represented by GANs and diffusion models~\cite{akter2024pix2pix,yu2024diffcast,shi2026wavec2r}. 
Despite these advances, most existing methods learn the radiance-to-reflectivity mapping mainly from local spectral and spatial cues, such as brightness-temperature gradients, cloud-top textures, and pixel-level patterns (Fig.~\ref{fig:motivation}a). 
Such cues are informative but incomplete, because satellite channels provide only partial cloud-top observations, whereas radar precipitation is governed by latent scene-level states, including cloud microphysics, convective organization, storm maturity, vertical development, and environmental moisture. 
This information gap leads to structural non-uniqueness: visually similar satellite signatures may correspond to meteorologically different precipitation states. 
In practice, this ambiguity manifests as cirrus false alarms when thin cirrus and deep convective anvils share similar cloud-top appearances, and as weakened or underestimated convective cores when pixel-level optimization averages over multiple plausible radar solutions. 
These limitations suggest that accurate S2R retrieval requires constraints beyond local visual cues, namely scene-level meteorological context that can narrow the plausible radar solution space.

Structured meteorological semantics provide a natural complement to local spectral cues for resolving the ambiguity of S2R retrieval. 
Recent multimodal weather foundation models have shown that language can encode global meteorological context and benefit weather understanding or generation tasks~\cite{nguyen2023climax,chen2023fengwu,zhao2024weathergfm,zhou2025omni}. 
For example, Omni-Weather introduces language-based task instructions and reasoning prompts to guide weather modeling, indicating that semantic information can provide useful high-level cues beyond raw meteorological fields (Fig.~\ref{fig:motivation}b). 
Nevertheless, existing uses of language in weather modeling remain largely auxiliary and open-loop: semantics are typically provided as fixed prompts, task instructions, or external annotations, without being explicitly tied to downstream retrieval accuracy. 
This is insufficient for S2R retrieval, where one-shot semantic cues may weaken along the radar generation trajectory, and externally generated descriptions may be meteorologically plausible but misaligned with retrieval-oriented metrics. 
In addition, online VLM inference can be costly for resource-constrained deployment. 
These limitations call for a task-specific semantic mechanism that remains effective throughout generation and can be adapted according to downstream retrieval performance.

To address these challenges, we propose LangRetrieval, a language-guided S2R retrieval framework that establishes closed-loop optimization between structured meteorological semantics and radar retrieval accuracy (Fig.~\ref{fig:motivation}c). 
LangRetrieval addresses the above limitations through two complementary mechanisms: trajectory-level semantic conditioning and retrieval-driven semantic optimization.
First, \textit{Semantic Warm-up} injects structured meteorological semantics into a conditional flow matching (CFM) backbone through cross-attention conditioning, allowing scene-level meteorological context to guide velocity estimation throughout the generation trajectory instead of being used only once at the input. 
Second, \textit{Self-Evolving Semantic Optimization} introduces a lightweight attribute policy that is initialized from offline vision-language model (VLM) annotations and subsequently refined via Group Relative Policy Optimization (GRPO) with multi-threshold Critical Success Index (CSI) rewards. 
This refinement moves semantic generation beyond merely imitating external annotations and aligns it with downstream retrieval skill, while avoiding online VLM inference at test time. 
Together, these designs transform structured meteorological semantics from passive auxiliary cues into CSI-aligned, retrieval-optimized representations for ill-posed S2R retrieval. 
As shown in Fig.~\ref{fig:motivation}d, LangRetrieval achieves a favorable quality-speed trade-off, maintaining strong perceptual quality while attaining up to $122.33\times$ lower inference latency than the diffusion baseline. 

The main contributions are summarized as follows:
\begin{itemize}
  \item We propose LangRetrieval, a language-guided S2R retrieval framework that establishes closed-loop optimization between meteorological semantics and retrieval accuracy, treating language as a learnable component directly optimized by task performance.
  
  \item We introduce attention-guided semantic conditioning within conditional flow matching, enabling continuous meteorological steering throughout the generation trajectory and addressing the global-context blindness of purely pixel-driven approaches.
  
  \item We develop a self-evolving attribute policy that combines supervised warm-up on VLM annotations with autonomous refinement via GRPO using retrieval-aware CSI rewards, achieving task-specific semantic optimization without VLM inference overhead at test time.
\end{itemize}
\section{Related Work}
\label{sec:related}

\subsection{Satellite-to-radar retrieval}
Satellite-to-radar (S2R) retrieval, which estimates ground-based radar reflectivity from geostationary satellite observations, is a critical capability for extending precipitation monitoring to radar-sparse regions. Deep learning approaches have substantially advanced this domain, from CNN-based UNet architectures~\cite{jin2023rda,zhao2024intelligent} to recent generative models. These include diffusion-based methods like DiffSR~\cite{yu2024diffcast, he2025diffsr}, which reconstructs composite radar reflectivity from satellite infrared and lightning inputs, and frequency-domain decomposition approaches like WaveC2R~\cite{shi2026wavec2r}. Furthermore, recent conditional flow matching approaches~\cite{tong2023conditional} have demonstrated superior computational efficiency and training stability over traditional diffusion models. With initial validation in weather nowcasting showing improvements in both inference speed and prediction accuracy~\cite{ribeiro2025flowcast}, flow-matching architectures have emerged as highly attractive backbones for weather generation tasks. However, despite these advances, existing S2R methods primarily operate on local spectral features-brightness temperature gradients, spatial patterns, pixel-level textures-without explicit modeling of the global meteorological semantics that govern precipitation intensity. Consequently, existing S2R retrievals have not yet leveraged the continuous, fine-grained guidance capabilities that flow matching enables to address these semantic shortcomings. 

\subsection{Language-guided weather models}
Recent advances in multimodal weather foundation models have explored incorporating language into weather understanding and generation pipelines~\cite{waqas2025artificial,varambally2025aquilon,li2025cllmate,han2025physics,shi2025deep,hong2026foundation}. Foundation models such as ClimaX~\cite{nguyen2023climax}, FengWu~\cite{chen2023fengwu}, and WeatherGFM~\cite{zhao2024weathergfm} leverage textual metadata as auxiliary conditioning signals, while Omni-Weather~\cite{zhou2025omni} demonstrates multi-task reasoning through language descriptions and chain-of-thought prompts. These models show that language can provide interpretability and multi-task flexibility in weather modeling.Yet these approaches treat language as a fixed, externally-provided signal-either auxiliary supervision during training or static inference prompts-without establishing mechanisms to optimize semantic content against downstream task objectives~\cite{10829821,11209564,11367371,11351323,11340758,11373262,11397268}. Critically, no closed-loop mechanism exists where semantic quality directly determines prediction accuracy: better descriptions do not systematically lead to better retrievals, because language remains decoupled from task-specific optimization. This represents a fundamental gap between language as auxiliary conditioning versus language as an actively-optimized component of the generation pipeline.

\subsection{Semantic optimization through reinforcement learning}
Reinforcement learning from human feedback (RLHF)~\cite{bai2022training} has become standard for aligning language models with human preferences~\cite{11198794,11154240,11140447,10888446,yan2025review}. More recent approaches such as Group Relative Policy Optimization (GRPO)~\cite{cai2025training} achieve comparable or superior performance with reduced computational overhead by replacing learned critics with group-relative baselines, as demonstrated in DeepSeek-R1~\cite{guo2025deepseek} for reasoning tasks.Moreover, practical deployment scenarios often involve hardware-limited environments where full-scale foundation models are prohibitive, necessitating lightweight yet semantically-aware retrieval pipelines.

In this paper, to our knowledge, this work is the first to apply RL-based semantic optimization with a purely physical reward signal (multi-threshold CSI) to a weather generation task. More fundamentally, we establish the first closed-loop mechanism between semantic content and prediction quality: a learned policy network generates task-specific meteorological descriptions that are directly optimized through downstream retrieval metrics, creating an explicit causal chain where semantic quality deterministically improves retrieval accuracy. This transforms language from a static auxiliary feature into a learnable, task-optimized component of the generation pipeline.

\begin{figure*}[t]
\centering
  \includegraphics[width=\textwidth]{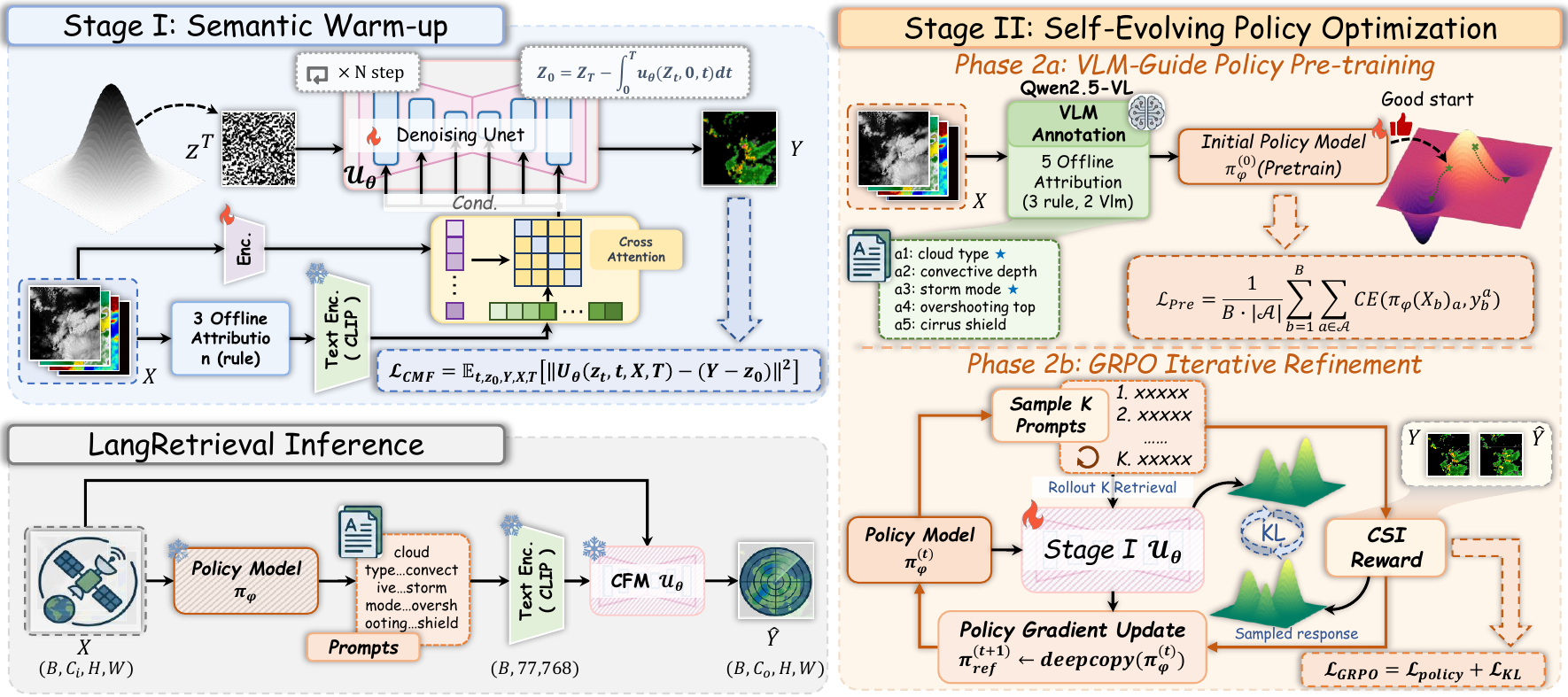}
\caption{Overview of the proposed LangRetrieval framework.
\textbf{Stage~I} (Semantic Warm-up): a CFM backbone $U_\theta$
transports noise to radar reflectivity via an ODE, conditioned on
meteorological semantic tokens via per-step cross-attention.
\textbf{Stage~II} (Self-Evolving Policy Optimization):
Phase~2a pre-trains policy $\pi_\phi$ on VLM-annotated attributes
via supervised cross-entropy; Phase~2b refines $\pi_\phi$ via
GRPO with multi-threshold CSI rewards and KL regularisation.
At inference, $\pi_\phi$ generates semantic prompts from satellite
input, which are CLIP-encoded and fed into $U_\theta$ to produce
the final VIL estimate.}
  \vspace{-15pt}
  \label{fig:framework}
\end{figure*}
\section{Methodology}
\label{sec:method}

\subsection{Problem Definition}
\label{sec:problem}

We formulate S2R retrieval as a conditional dense prediction problem from geostationary satellite observations to radar-based precipitation fields.
Given synchronous multi-spectral satellite observations
$X \in \mathbb{R}^{s\times H\times W}$, comprising visible imagery $X_{\rm vis}$ and dual-channel infrared imagery $X_{\rm ir069}$ and $X_{\rm ir107}$, where $s$ denotes the number of input channels, the goal is to generate the corresponding radar-based precipitation field
$\hat{Y} \in \mathbb{R}^{1 \times H \times W}$ that approximates the ground truth $Y \in \mathbb{R}^{1 \times H \times W}$:
\begin{equation}
\begin{aligned}
&\hat{Y} = f_\theta(X), \\
&\theta^* = \arg\min_\theta\,
  \mathbb{E}_{(X,Y)\sim\mathcal{D}}
  \bigl[\mathcal{L}_{\rm ret}(f_\theta(X),\,Y)\bigr],
\end{aligned}
\label{eq:s2r_problem}
\end{equation}
where $f_\theta$ denotes the retrieval model with learnable parameters $\theta$, $\mathcal{D}$ is the joint distribution of satellite--radar pairs, and $\mathcal{L}_{\rm ret}$ represents a task-driven retrieval objective.
In LangRetrieval, $f_\theta$ is instantiated as a language-guided generative pipeline in which structured meteorological semantics inferred from $X$ condition the radar-field generation process, as detailed in Section~\ref{sec:overview}.

\subsection{Preliminary of Conditional Flow Matching}
\label{sec:prelim}

Conditional flow matching (CFM)~\cite{lipman2022flow} learns a
time-dependent velocity field that transports samples from a source
distribution $\mathcal{N}(0,I)$ to a target data distribution via
an ODE, conditioned on auxiliary context. Given an initial noise sample $z_0 \sim \mathcal{N}(0,I)$ and target $Y \sim p_{\rm data}$,
CFM constructs a probability path by linear interpolation:
\begin{equation}
  z_t = (1-t)\,z_0 + t\,Y,
  \quad t \in [0,1],
  \label{eq:cfm_path}
\end{equation}
where $z_t$ is the interpolated state at time $t$.

The conditional vector field along this path has an analytic closed-form expression: $u_t(z_t \mid Y) = Y - z_0$.
A velocity estimator $U_\theta$ is trained to regress this target by minimizing the following objective:
\begin{equation}
  \mathcal{L}_{\rm CFM}(\theta) =
  \mathbb{E}_{t,\,z_0,\,Y}
  \bigl[\,
    \|\,U_\theta(z_t,\,t) - (Y - z_0)\,\|^2
  \,\bigr],
  \label{eq:cfm_base}
\end{equation}
where $U_\theta(z_t,t)$ is the predicted velocity field at state $z_t$ and time $t$.
At inference, starting from $z_0 \sim \mathcal{N}(0,I)$, we obtain $\hat{Y} = z_1$ by integrating $\mathrm{d}z_t/\mathrm{d}t = U_\theta(z_t,t)$ using a fixed-step Euler solver, requiring substantially fewer function evaluations than score-based diffusion models. In our LangRetrieval framework, we extend
Eqs.~\eqref{eq:cfm_path}--\eqref{eq:cfm_base} to the conditional
setting by augmenting the velocity estimator with satellite observations $X$ and structured semantic tokens $T\in\mathbb{R}^{d}$, i.e., $U_\theta(z_t, t, X, T)$ (see Section~\ref{sec:stage1} for details).

\subsection{Framework Overview}
\label{sec:overview}
LangRetrieval decomposes the end-to-end mapping $X \mapsto \hat{Y}$ into a three-component closed-loop pipeline illustrated in Fig.~\ref{fig:framework}:
\begin{equation}
  \underbrace{\bm{a} = \pi_\phi(X)}_{\text{attribute policy}}
  \;\longrightarrow\;
  \underbrace{T = \mathcal{E}_\psi(s(\bm{a}))}_{\text{text encoder}}
  \;\longrightarrow\;
  \underbrace{\hat{Y} =
    \mathrm{ODE}[U_\theta(\,\cdot\,,\,\cdot\,,X,T)]}_{\text{retrieval backbone}},
  \label{eq:pipeline}
\end{equation}
where $\bm{a}=(a_1,\ldots,a_5)$ is a five-dimensional meteorological attribute vector, $\pi_\phi$ denotes the attribute policy network, $s(\cdot)$ is a text template function, $\mathcal{E}_\psi$ is a frozen CLIP text encoder, and $U_\theta$ is the CFM retrieval backbone conditioned on both $X$ and $T$ as defined in Eq.~\eqref{eq:stage1_loss}.
The mutual dependency between $\pi_\phi$ and $U_\theta$ renders naive end-to-end optimization intractable, necessitating the sequential two-stage bootstrap strategy summarized in Table~\ref{tab:training_stages}.

\textbf{Stage~I: Semantic Warm-Up.}
Resolving the ill-posed radiance-to-reflectivity mapping 
requires semantics to persistently constrain the generation trajectory at every ODE step rather than being consumed once at the input boundary.
To activate this capability, Stage~I trains $U_\theta$ in 
isolation to internalize structured meteorological semantics into its CFM backbone via per-step cross-attention conditioning.
The three pixel-verifiable attributes $\{a_2, a_4, a_5\}$ are deterministically derived from satellite input fields and serve as reliable conditioning signals at this stage, while $a_1$ (cloud type) and $a_3$ (storm mode), which require holistic morphological reasoning beyond pixel-level statistics, are held as \texttt{Unknown}.
Under this partial but objective conditioning, $U_\theta$ learns to modulate its generation trajectory in response to semantic tokens at each integration step, establishing the semantic-conditioned generation space that serves as a 
functional reward environment for Stage~II.

\textbf{Stage~II: Self-Evolving Policy Optimization.}
While Stage~I equips $U_\theta$ with semantic responsiveness, the semantics are never optimized for retrieval: $a_1$ and $a_3$ remain absent, static rules cannot capture real-world meteorological variability, and no offline supervision signal can anticipate which attribute combinations maximize CSI within the evolving generation space of $U_\theta$.
Closed-loop optimization between semantic generation and 
retrieval feedback is therefore indispensable.
To this end, we introduce a lightweight $\pi_\phi$ that 
predicts all five attributes directly from $X$ with a single forward pass, eliminating VLM dependency at test time while enabling scene-adaptive semantic generation.
$\pi_\phi$ is first pretrained on Qwen-VL annotations via 
supervised cross-entropy to establish a coherent semantic 
prior and prevent reward collapse.
Subsequently, $\pi_\phi$ and $U_\theta$ are jointly optimized via GRPO under multi-threshold CSI rewards, realigning the optimization target of $\pi_\phi$ from descriptive fidelity toward retrieval accuracy.
\begin{table}[t]
  \centering
  \caption{Parameter update schedule across training stages.
           \Updated: updated; \Frozen: frozen.}
  \label{tab:training_stages}
  \setlength{\tabcolsep}{14pt}
  \begin{tabular}{lccc}
    \toprule
    Stage & $U_\theta$ & $\mathcal{E}_\psi$ & $\pi_\phi$ \\
    \midrule
    Stage~I: Semantic Warm-Up     & \Updated & \Frozen   & \Frozen   \\
    Phase~2a: Policy Pre-training & \Frozen   & \Frozen   & \Updated \\
    Phase~2b: GRPO Fine-tuning    & \Updated   & \Frozen   & \Updated \\
    \bottomrule
  \end{tabular}
  \vspace{-4mm}
\end{table}

\subsection{Stage~I: Semantic Warm-Up}
\label{sec:stage1}

This stage establishes a semantically-warmed retrieval backbone $U_\theta$ to mitigate the underdetermined radiance-to-reflectivity mapping by incorporating structured meteorological semantics into the generation process.
We instantiate $U_\theta$ as a CFM-based backbone, whose 
deterministic ODE formulation delivers high-fidelity retrieval with substantially fewer function evaluations than stochastic diffusion models~\cite{feng2025perceptually,ribeiro2025flowcast}, 
making it well-suited for resource-constrained deployment.
Since the generation unfolds as a continuous ODE trajectory, semantic tokens $T$ are injected at every integration step via cross-attention rather than applied once at the input boundary, ensuring that global meteorological context persistently constrains the generation trajectory throughout.
$U_\theta$ follows a UNet architecture with base channel 
width 64, channel multipliers $\{1,2,4,8\}$, and 
time-conditioned ResBlocks (see supplementary material 
for full architecture details).

\paragraph{Partial Semantic Prompt}
Since $\pi_\phi$ is not yet available in this stage, only the three pixel-verifiable attributes $\{a_2,a_4,a_5\}$ enter the prompt, while the remaining attributes $a_1$ and $a_3$ are set to the placeholder \texttt{Unknown}. The resulting text $s(\bm{a})$ is encoded by the frozen CLIP text encoder $\mathcal{E}_\psi$ (CLIP ViT-L/14, $d_{\rm text}=768$), producing semantic tokens $T \in \mathbb{R}^{L \times d_{\rm text}}$.

\paragraph{Semantic Cross-Attention Injection}
Satellite visual features encode \emph{where}
precipitation is likely to occur, while the semantic tokens characterize \emph{what kind} of meteorological structure is present. To exploit this complementarity, we assign $V_t$ as the query source, enabling each spatial location to actively attend to and retrieve relevant global scene context from $T$, rather than receiving it via passive broadcast.
Concretely, at two intermediate UNet resolutions
($32{\times}32$ and $64{\times}64$), the flattened visual
feature map $V_t \in \mathbb{R}^{N \times C}$ is projected into queries, where $N$ and $C$ denote the number of spatial locations and channels. The semantic tokens $T$ are projected to keys and values.
The resulting attention output is then added via a residual connection and followed by layer normalization to produce the updated feature $\bar{V}_t$. This process is formally defined as:
\begin{equation}
  \bar{V}_t = \mathrm{LN}\bigl(
    \mathrm{Attn}(V_tW_Q,\;TW_K,\;TW_V) + V_t
  \bigr),
  \label{eq:attn}
\end{equation}
where $\mathrm{Attn}(Q,K,V)=\mathrm{Softmax}(QK^\top\!/\sqrt{d_k})\,V$. $W_Q \in \mathbb{R}^{C \times d_k}$, $W_K, W_V \in \mathbb{R}^{d_{\rm text} \times d_k}$ are learnable projection matrices, and $\mathrm{LN}(\cdot)$ denotes layer normalization. This mechanism injects global meteorological context into the generation process at every ODE step.

\paragraph{Training Objective}
Extending Eq.~\eqref{eq:cfm_base} to the conditional setting,
we sample the timestep $t$ from a logit-normal distribution
$\mathrm{logit}(t)\!\sim\!\mathcal{N}(-0.4,\,1.0)$ to
up-weight informative intermediate steps. We can formulate the training objective as:
\begin{equation}
  \mathcal{L}_{\rm CFM}(\theta)
  = \mathbb{E}_{t,\,z_0,\,Y,\,X,\,T}
    \Bigl[\,
      \bigl\|\,
        U_\theta(z_t,\,t,\,X,\,T) - (Y - z_0)
      \,\bigr\|^2
    \,\Bigr],
  \label{eq:stage1_loss}
\end{equation}
where $U_\theta(z_t,t,X,T)$ is the predicted velocity field conditioned on satellite input $X$ and semantic tokens $T$.
Although this objective yields strong structural fidelity,
the semantic tokens $T$ are produced by a large frozen
vision-language encoder $\mathcal{E}_\psi$, which is
prohibitively expensive for edge meteorological platforms
with tight memory and compute constraints.
Stage~II therefore replaces this VLM dependency with a
lightweight self-evolving policy $\pi_\phi$, shifting the
optimization focus from generation fidelity to retrieval accuracy.

\subsection{Stage~II: Self-Evolving Policy Optimization}
\label{sec:stage2}
To further improve retrieval accuracy by aligning semantic 
generation directly with CSI objectives, and to enable 
practical deployment in resource-constrained environments 
where online VLM inference is unavailable, Stage~II 
introduces a \emph{self-evolving} attribute policy $\pi_\phi$ instantiated as a lightweight CNN.
$\pi_\phi$ first imitates VLM-derived annotations via 
supervised pre-training (Phase~2a) to establish a coherent 
semantic prior, then refines its predictions through 
closed-loop reward optimization (Phase~2b), with both 
$\pi_\phi$ and $U_\theta$ jointly optimized under 
multi-threshold CSI rewards to evolve beyond the VLM 
teacher toward retrieval accuracy.
Formally, $\pi_\phi$ maps satellite observation $X$ to 
categorical distributions over the five attributes:
\begin{equation}
  p_\phi(a_i \mid X) = \mathrm{softmax}\bigl(h_i(f(X))\bigr),
  \quad i = 1, \ldots, 5,
  \label{eq:policy}
\end{equation}
where $f$ denotes a convolutional encoder followed by global average pooling, $h_i$ is the linear prediction head for the $i$-th attribute, and $p_\phi(a_i \mid X)$ represents the predicted categorical distribution for attribute $a_i$.
Directly optimizing $\pi_\phi$ from random initialization using only task rewards is highly unstable. Therefore, training is conducted in two sequential phases.

\subsubsection{Phase~2a: Semantic-Guide Policy Pre-training}
\label{sec:phase2a}

\paragraph{Five-Attribute Semantic Taxonomy.}
We describe each satellite observation with a structured 
five-attribute vector $P = (a_1, a_2, a_3, a_4, a_5)$ 
that encodes the global cloud properties governing the 
radiance-precipitation relationship, comprising three 
cloud-structure attributes $(a_1, a_2, a_3)$ and two 
convective indicators $(a_4, a_5)$:
\begin{itemize}
  \item $a_1$ (\texttt{cloud\_type}): Cb, Ci, Ns, Sc;
  \item $a_2$ (\texttt{convective\_depth}): Deep, Moderate, Shallow, None;
  \item $a_3$ (\texttt{storm\_mode}): Supercell, Squall-line, MCS, Isolated, Stratiform;
  \item $a_4$ (\texttt{overshooting\_top}): Present, Absent;
  \item $a_5$ (\texttt{cirrus\_shield}): Dominant, Partial, Absent.
\end{itemize}

Attributes $a_2,a_4,a_5$ are deterministically derived from the IR 10.7\,\textmu m brightness temperature field, requiring no external model during training or inference (see rules in supplementary material).
In contrast, $a_1$ and $a_3$ require holistic visual reasoning over cloud morphology beyond pixel-level statistics, and we therefore obtain them via offline annotations from Qwen-VL on the training set.
At inference stage, all five attributes are predicted by
the policy network $\pi_\phi$. The predicted attributes are concatenated into a structured textual prompt $s$ and encoded
by a frozen CLIP ViT-L/14 text encoder to produce semantic tokens $T = \mathcal{E}_\psi(s) \in \mathbb{R}^{L \times d_{\rm text}}$.

\paragraph{Pre-training Objective.}
The policy $\pi_\phi$ is pretrained using a cross-entropy objective on the Qwen-VL annotations:
\begin{equation}
  \mathcal{L}_{\rm pre}(\phi)
  = \frac{1}{|B|\cdot|A|}
    \sum_{b \in B}\sum_{i \in A}
    \mathrm{CE}\!\bigl(p_\phi(a_i \mid X_b),\,
                      \hat{a}_i^{(b)}\bigr),
  \label{eq:pretrain_loss}
\end{equation}
where $B$ is the mini-batch, $A = \{1,\ldots,5\}$ is the attribute index set, $\hat{a}_i^{(b)}$ is the pseudo-ground-truth label for attribute $i$ in sample $b$, and $\mathrm{CE}$ denotes cross-entropy loss.
This stage yields an initialization $\pi_\phi^{(0)}$ that encapsulates VLM-derived semantic priors, providing a stable foundation for subsequent reward-driven optimization.

\subsubsection{Phase~2b: GRPO Iterative Refinement}
\label{sec:phase2b}

Phase~2b drives the self-evolution of $\pi_\phi$ and $U_\theta$ beyond the VLM imitation ceiling: both modules are jointly optimized so that $\pi_\phi$ autonomously discovers which attribute vectors maximize downstream CSI while $U_\theta$ simultaneously adapts to leverage the evolving semantics.
This closed-loop optimization is conducted over $N$ \textit{Iterative Refinement Rounds} (IRRs) using GRPO. The overall procedure is formalized in Algorithm~\ref{alg:grpo}, with each training step within a round detailed as follows:

\paragraph{Group Sampling}
For each input sample $X_b$, we draw $K$ candidate attribute vectors $\{\bm{a}_b^{(k)}\}_{k=1}^K$ from $\pi_\phi(\cdot \mid X_b)$ using temperature $T_s$, and record their joint log-probabilities $\ell_b^{(k)} = \log\pi_\phi(\bm{a}_b^{(k)} \mid X_b)$,
where $b$ indexes samples in the batch and $k$ indexes the candidates within each group.

\paragraph{Backbone Rollout}
Each attribute vector $\bm{a}_b^{(k)}$ is encoded via the frozen CLIP encoder $\mathcal{E}_\psi$ to obtain semantic tokens $T_b^{(k)}$. $U_\theta$ is then integrated jointly with $\pi_\phi$ to produce the radar precipitation estimate:
\begin{equation}
  \hat{Y}_b^{(k)}
  = \mathrm{EulerODE}\!\bigl(U_\theta,\, X_b,\, T_b^{(k)}\bigr),
  \label{eq:rollout}
\end{equation}
where $\hat{Y}_b^{(k)} \in \mathbb{R}^{1 \times H \times W}$ denotes the predicted radar reflectivity for candidate $k$ of sample $b$.

\paragraph{Reward and Advantage Estimation}
We define an objective, annotation-free reward based on the multi-threshold Critical Success Index (CSI):
\begin{equation}
  r(\hat{Y},\,Y)
  = \sum_{\tau \in \mathcal{T}} w_\tau \cdot
    \frac{\mathrm{TP}_\tau}
         {\mathrm{TP}_\tau + \mathrm{FP}_\tau + \mathrm{FN}_\tau},
  \label{eq:reward}
\end{equation}
where $\mathcal{T}$\,dBZ is the set of reflectivity thresholds, $w_\tau$ are corresponding weights. $\mathrm{TP}_\tau$, $\mathrm{FP}_\tau$, $\mathrm{FN}_\tau$ denote true positives, false positives, and false negatives at threshold $\tau$. The largest weight is assigned to the heavy-precipitation regime, which is operationally most consequential. The group-relative advantage is computed as:
\begin{equation}
  A_b^{(k)} = r_b^{(k)} - \bar{r}_b,
  \qquad
  \bar{r}_b = \frac{1}{K}\sum_{k=1}^{K} r_b^{(k)},
  \label{eq:advantage}
\end{equation}
where $r_b^{(k)} = r(\hat{Y}_b^{(k)}, Y_b)$ is the reward for candidate $k$ of sample $b$, $\bar{r}_b$ is the group mean reward serving as the baseline. 

\paragraph{Policy Update}
The GRPO objective combines an advantage-weighted policy gradient with KL regularization:
\begin{equation}
\begin{split}
  \mathcal{L}_{\rm GRPO}(\phi)
  &= -\frac{1}{BK}\sum_{b,k} A_b^{(k)} \cdot \ell_b^{(k)} \\
  &\quad + \beta \sum_{i=1}^{5}
      \mathrm{KL}\!\bigl(
        p_\phi(a_i \mid X) \,\|\, p_{\rm ref}(a_i \mid X)
      \bigr),
\end{split}
  \label{eq:grpo_loss}
\end{equation}
where $\beta$ is the KL penalty coefficient, and $p_{\rm ref}$ denotes the reference policy distribution. 
At the end of each IRR, the reference policy is updated via:
$\pi_{\rm ref} \leftarrow \mathrm{deepcopy}(\pi_\phi)$, so that subsequent rounds explore from the best-performing policy obtained so far. This progressive update enables continual self-improvement across rounds, while the within-round KL regularization prevents collapse of the attribute distribution.

\begin{algorithm}[t]
\caption{Self-Evolving Policy Optimisation via
         GRPO with Progressive Reference Reset}
\label{alg:grpo}
\begin{algorithmic}[1]
\REQUIRE $\pi_\phi$, $U_\theta$, frozen $\mathcal{E}_\psi$,
         $\mathcal{D}$, rounds $N$, group size $K$, KL weight $\beta$
\STATE $\pi_{\rm ref} \leftarrow \pi_\phi$
\FOR{$n = 1, \ldots, N$}
  \FOR{each $(X_b, Y_b) \in \mathcal{D}$}
    \STATE Sample $\{\bm{a}_b^{(k)},\,\ell_b^{(k)}\}_{k=1}^K$
           from $\pi_\phi(\cdot\mid X_b)$
    \STATE $\hat{Y}_b^{(k)} \leftarrow
           \mathrm{EulerODE}(U_\theta,\,X_b,\,
           \mathcal{E}_\psi(\bm{a}_b^{(k)}))$
           \hfill\eqref{eq:rollout}
    \STATE Compute $r_b^{(k)}$, $A_b^{(k)}$
           \hfill\eqref{eq:reward}\eqref{eq:advantage}
    \STATE $\phi \leftarrow \phi
           - \alpha\nabla_\phi\mathcal{L}_{\rm GRPO}$
           \hfill\eqref{eq:grpo_loss}
    \STATE $\theta \leftarrow \theta
           - \alpha\nabla_\theta\mathcal{L}_{\rm GRPO}$
           \hfill\textit{(joint backbone update)}
  \ENDFOR
  \STATE $\pi_{\rm ref} \leftarrow \mathrm{deepcopy}(\pi_\phi)$
         \hfill\textit{(progressive reference reset)}
\ENDFOR
\ENSURE Optimised $\pi_\phi$, $U_\theta$
\end{algorithmic}
\end{algorithm}
\subsection{Inference}
\label{sec:inference}
At test time, all three components are frozen, and the pipeline requires only the satellite observation $X$ as input. The lightweight policy $\pi_\phi$ performs a single forward pass to predict the most likely attribute vector $\hat{\bm{a}} = \arg\max_{\bm{a}}\,\pi_\phi(\bm{a} \mid X)$,
which is then assembled into a structured text prompt and encoded by the frozen CLIP encoder to produce semantic tokens $T$. Conditioned on $(X, T)$, the backbone $U_\theta$ integrates the velocity field from $z_0 \sim \mathcal{N}(0,I)$ to produce the final radar estimate $\hat{Y} = z_1$ via a fixed-step Euler solver. Notably, the inference process requires no VLM calls, no iterative refinement, and no additional inputs beyond satellite observation $X$.

\section{Experiments}
\label{sec:experiments}

\subsection{Experimental Setup}

\begin{figure}
    \centering
    \includegraphics[width=0.5\textwidth]{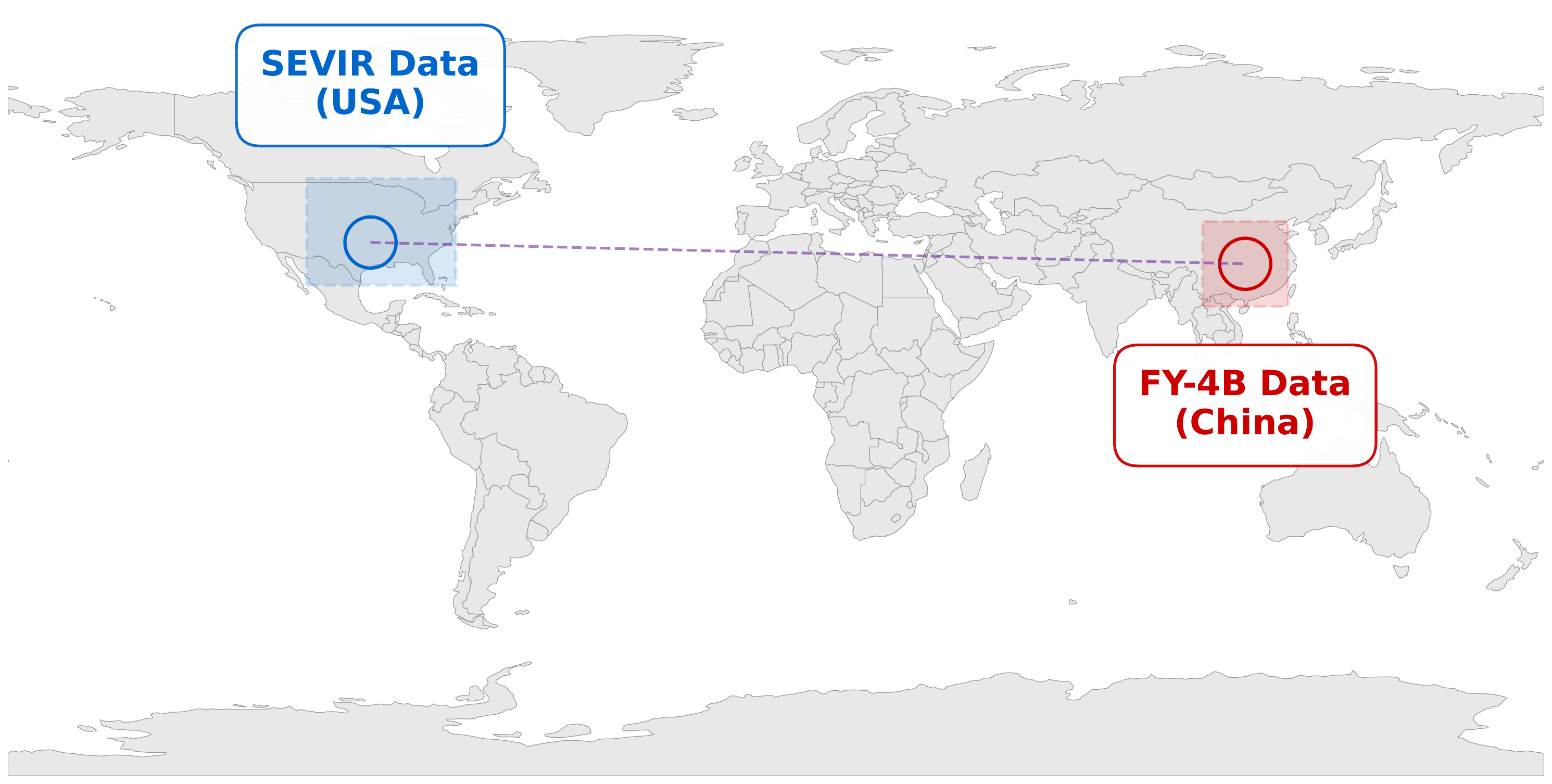}
    \vspace{-3mm}
    \caption{Geographic distribution of evaluation datasets. The SEVIR dataset covers the continental United States (blue region), while the Southeast China FY-4B dataset spans 100$^{\circ}$--120$^{\circ}$E and 20$^{\circ}$--40$^{\circ}$N (red region). The dashed line indicates cross-domain validation between the two datasets.}
    \label{fig:study_area}
\end{figure}

\noindent\textbf{Datasets.} 
We conduct experiments on two precipitation forecasting datasets (Fig.~\ref{fig:study_area}). The \textit{SEVIR} \footnote{\url{https://registry.opendata.aws/sevir/}} dataset~\cite{veillette2020sevir} covers the United States with over 20,000 storm events (2017-2019) at 1 km spatial resolution ($384 \times 384$ pixels) and 5-minute temporal intervals, comprising visible, IR069, IR107, Vertically Integrated Liquid (VIL), and lightning observations. The \textit{Southeast China FY-4B\footnote{\url{https://satellite.nsmc.org.cn/portalsite/Data/DataView.aspx?SatelliteType=1\&SatelliteCode=FY4B}} dataset} spans 100$^{\circ}$--120$^{\circ}$E, 20$^{\circ}$--40$^{\circ}$N (2023) with paired satellite observations (VIS, WV, IR) and radar reflectivity at 4 km resolution ($500 \times 500$ pixels) and 30-minute intervals. Following standard practice, all images are reshaped to $128 \times 128$ for consistent evaluation. Dataset splits are provided in Table~\ref{tab:datasets}.

\begin{table}[h]
\centering
\caption{Summary of datasets used for evaluation.}
\label{tab:datasets}
\resizebox{0.5\textwidth}{!}{
\begin{tabular}{lccccc}
\toprule
\textbf{Dataset} & $N_{\text{train}}$ & $N_{\text{val}}$ & $N_{\text{test}}$ & \textbf{Resolution} & \textbf{Temporal}\\
\midrule
SEVIR & 12,331 & 4,389 & 4,943 & 1 km & 5 min \\
FY4B-SEChina & 4,568 & 1,144 & 1,114 & 4 km & 30 min \\
\bottomrule
\end{tabular}}
\end{table}

\begin{table*}[h]
\normalsize
\centering
\caption{
    Quantitative comparison with state-of-the-art methods on \textbf{SEVIR} test set.
    CSI, HSS$\uparrow$: higher is better; LPIPS, MAE$\downarrow$: lower is better.
    \textbf{Bold}: best; \second{underline}: second best.
}
\label{tab:sota}
\resizebox{1\textwidth}{!}{
\setlength{\tabcolsep}{3.5pt}
\renewcommand{\arraystretch}{1.2}
\begin{tabular}{@{}p{0.8cm}c|cccccc|cccccc|ccc@{}}
\toprule
\multirow{3}{*}{\centering\textbf{Type}}
& \multirow{3}{*}{\textbf{Model}}
& \multicolumn{6}{c}{\textbf{CSI}$\uparrow$}
& \multicolumn{6}{c}{\textbf{HSS}$\uparrow$}
& \multicolumn{2}{c}{\textbf{Perceptual quality}} \\
\cmidrule(lr){3-8} \cmidrule(lr){9-14} \cmidrule(lr){15-16}
& & @74 & @133 & @160 & @181 & @219 &Avg. & @74 & @133 & @160 & @181 & @219 &Avg.&MAE$\downarrow$&LPIPS$\downarrow$\\
\midrule\midrule

\multirow{9}{*}{\centering\rotatebox{90}{\textbf{SEVIR}}}
& AA-TransUnet~\cite{yang2022aa} 
&0.506 & 0.329& 0.317 & 0.262&0.095&0.301
&0.638 &0.483 &\second{0.476} &0.412& 0.174&0.437
& \second{9.76} & 0.318\\
& Earthformer~\cite{gao2022earthformer} 
&\second{0.524} &0.321 &0.304  &0.249 &0.069 &0.293
&\second{0.653} &0.474  &0.461 &0.395 &0.129 &0.422
& \best{9.71} & 0.360\\
& Smaat-Unet~\cite{trebing2021smaat} 
&0.516 &\second{0.324} &0.313  &0.256 & 0.055 &0.292
&0.646 & 0.478 &0.471 &0.404 &0.105 &0.420
& 9.77 & 0.362\\
& Diffcast~\cite{yu2024diffcast} 
&0.501 &0.322 &\second{0.321}&\second{0.282} &\second{0.126} &\second{0.310}
&0.629 &0.473 &0.479&\best{0.436}&\second{0.223} &\second{0.448}
& 11.42 & \second{0.236} \\
& Pix2Pix~\cite{akter2024generative} 
&0.418 &0.259 & 0.269 & 0.241& 0.014&0.240
&0.547 &\second{0.397} &0.417 &0.384 & 0.028 &0.355
& 11.77 & 0.349\\
& MeanFlow~\cite{geng2025mean} 
&0.425 & 0.215& 0.177 &0.141 &0.044 &0.200
&0.550 & 0.333&0.290 & 0.242 &0.084 &0.299
& 14.20 & 0.300\\
\cmidrule(lr){2-16}

\rowcolor{orange}
& LangRetrieval (Ours)
&\best{0.536} &\best{0.405} &\best{0.369} &\best{0.307} &\best{0.185}&\best{0.360}
&\second{0.645} &\best{0.537} &\best{0.501} &\second{0.425} &\best{0.228} &\best{0.467}
&10.44 &\best{0.202}\\
\rowcolor{gray!20}
& \textcolor{red}{\textit{Improv. vs 2\textsuperscript{nd}}} 
& \textcolor{red}{+2.3\%} & \textcolor{red}{+24.7\%} & \textcolor{red}{+14.9\%} & \textcolor{red}{+8.9\%} & \textcolor{red}{+46.8\%} & \textcolor{red}{+16.1\%}
& \textcolor{red}{--} & \textcolor{red}{+11.2\%} & \textcolor{red}{+4.6\%} & \textcolor{red}{--} & \textcolor{red}{+2.2\%} & \textcolor{red}{+4.2\%}
& \textcolor{red}{--} & \textcolor{red}{+14.4\%}\\

\rowcolor{gray!20}
& \textcolor{red}{\textit{Improv. vs MeanFlow}} 
& \textcolor{red}{+20.7\%} & \textcolor{red}{+69.8\%} & \textcolor{red}{+91.0\%} & \textcolor{red}{+102.1\%} & \textcolor{red}{+234.1\%} & \textcolor{red}{+64.5\%}
& \textcolor{red}{+15.6\%} & \textcolor{red}{+53.5\%} & \textcolor{red}{+67.6\%} & \textcolor{red}{+75.2\%} & \textcolor{red}{+170.2\%} & \textcolor{red}{+52.5\%}
& \textcolor{red}{+21.9\%} & \textcolor{red}{+38.7\%}\\

\bottomrule
\end{tabular}}
\end{table*}

\begin{table*}[h]
\normalsize
\centering
\caption{
    Quantitative comparison with state-of-the-art methods on 
    \textbf{Southeast China FY-4B} test set.
    CSI, HSS$\uparrow$: higher is better; FAR$\downarrow$: lower is better.
    \textbf{Bold}: best; \second{underline}: second best.
}
\label{tab:sota_fy}
\resizebox{1\textwidth}{!}{
\setlength{\tabcolsep}{3.5pt}
\renewcommand{\arraystretch}{1.2}
\begin{tabular}{@{}p{0.8cm}c|ccccc|ccccc|cccccc@{}}
\toprule
\multirow{3}{*}{\centering\textbf{Type}}
& \multirow{3}{*}{\textbf{Model}}
& \multicolumn{5}{c}{\textbf{CSI}$\uparrow$}
& \multicolumn{5}{c}{\textbf{HSS}$\uparrow$}
& \multicolumn{5}{c}{\textbf{FAR}$\downarrow$} \\
\cmidrule(lr){3-7} \cmidrule(lr){8-12} \cmidrule(lr){13-17}
& & @10 & @20 & @25 & @30 & @35 & @10 & @20 & @25 & @30 & @35 & @10 & @20 & @25 & @30 & @35 \\
\midrule\midrule

\multirow{9}{*}{\centering\rotatebox{90}{\textbf{FY-4B}}}

& AA-TransUnet~\cite{yang2022aa}
& 0.323 & 0.234 & 0.186 & 0.130 &0.075
& 0.425 & 0.347 & 0.297 & 0.223 &0.084
& 0.350 & 0.430 & 0.480 & 0.518 &0.606\\

& Earthformer~\cite{gao2022earthformer} 
& \second{0.346} & 0.264 & 0.215 & 0.145 &0.060
& 0.454 & 0.387 & 0.336 & 0.245 &0.066
& 0.319 & 0.386 & 0.442 & 0.497 &0.603\\

& Smaat-Unet~\cite{trebing2021smaat}
& 0.253 & 0.203 & 0.164 & 0.118 &0.070
& 0.327 & 0.294 & 0.257 & 0.199 &0.092
& 0.480 & 0.591 & 0.648 & 0.701 &0.770\\

& Diffcast~\cite{yu2024diffcast}
& 0.345 & 0.263 & \second{0.227} & 0.174 &0.106
& \second{0.457} & \second{0.389} & 0.355 & 0.289 &0.120
& \best{0.270} & \second{0.316} & \second{0.373} & \second{0.430} &\best{0.521} \\

& Pix2Pix~\cite{akter2024generative}
& 0.324 & \second{0.267} & 0.233 & \second{0.189} &\second{0.132}
& 0.422 & 0.386 & \second{0.357} & \second{0.305} &\second{0.198}
& 0.386 & 0.459 & 0.523 & 0.608 &0.712\\

& MeanFlow~\cite{geng2025mean} 
& 0.307 & 0.241 & 0.201 & 0.148 &0.092
& 0.407 & 0.358 & 0.316 & 0.249 &0.128
& 0.395 & 0.475 & 0.555 & 0.639 &0.752\\

\cmidrule(lr){2-17}
\rowcolor{orange}
& LangRetrieval (Ours)
& \best{0.381} & \best{0.309} & \best{0.282} & \best{0.228} &\best{0.142}
& \best{0.487} & \best{0.435} & \best{0.413} & \best{0.349} &\best{0.233}
& \second{0.275} & \best{0.303} & \best{0.349} & \best{0.412} &\second{0.558}\\

\rowcolor{gray!20}
& \textcolor{red}{\textit{Improv. vs 2\textsuperscript{nd}}}
& \textcolor{red}{+10.1\%}  & \textcolor{red}{+15.7\%} & \textcolor{red}{+24.2\%} & \textcolor{red}{+20.6\%}&\textcolor{red}{+7.6\%}
& \textcolor{red}{+6.6\%}  & \textcolor{red}{+11.8\%} & \textcolor{red}{+15.7\%} & \textcolor{red}{+14.4\%}&\textcolor{red}{+17.7\%}
& \textcolor{red}{--}      & \textcolor{red}{+4.1\%}  & \textcolor{red}{+6.4\%}  & \textcolor{red}{+4.2\%} &\textcolor{red}{--}\\

\rowcolor{gray!20}
& \textcolor{red}{\textit{Improv. vs MeanFlow}}
& \textcolor{red}{+24.1\%} & \textcolor{red}{+28.2\%} & \textcolor{red}{+40.3\%} & \textcolor{red}{+54.1\%} &\textcolor{red}{+54.3\%}
& \textcolor{red}{+19.7\%} & \textcolor{red}{+21.5\%} & \textcolor{red}{+30.7\%} & \textcolor{red}{+40.2\%} &\textcolor{red}{+81.9\%}
& \textcolor{red}{+30.4\%} & \textcolor{red}{+36.2\%} & \textcolor{red}{+37.1\%} & \textcolor{red}{+35.5\%} &\textcolor{red}{+25.8\%}\\

\bottomrule
\end{tabular}}
\end{table*}

\begin{table}[t]
\small
\centering
\caption{
    Perceptual quality and inference speed comparison on
    \textbf{Southeast China FY-4B} benchmarks.
    SSIM, PSNR$\uparrow$: higher is better; MAE, LPIPS$\downarrow$: lower is better.
}
\label{tab:sota_perceptual}
\resizebox{0.5\textwidth}{!}{
\setlength{\tabcolsep}{3pt}
\renewcommand{\arraystretch}{1.2}
\begin{tabular}{@{}cccccccc}
\toprule
\multirow{2}{*}{\textbf{Model}}
& \multicolumn{7}{c}{\textbf{FY-4B}}\\
\cmidrule(lr){2-8}
& SSIM$\uparrow$ & PSNR$\uparrow$ & MAE$\downarrow$ & LPIPS$\downarrow$& Avg.CSI$\uparrow$& Avg.HSS$\uparrow$& Avg.FAR$\downarrow$\\
\midrule\midrule
AA-TransUnet~\cite{yang2022aa}
  & 0.539 & 20.49 & 2.980 & 0.280 & 0.189 & 0.275 & 0.476 \\
Earthformer~\cite{gao2022earthformer} 
  & 0.555 & 20.86 & 2.864 & 0.338 & 0.206 & 0.297 & 0.449 \\
Smaat-Unet~\cite{trebing2021smaat}
  & 0.501 & 19.18 & 3.498 & 0.333 &0.161 & 0.233 & 0.638  \\
Diffcast~\cite{yu2024diffcast}
  & \second{0.590} & \second{21.23} & \second{2.681} & 0.274 & 0.223 & 0.322 & \second{0.380} \\
Pix2Pix~\cite{akter2024generative}
  & 0.523 & 19.88 & 2.979 & 0.257 & \second{0.229}& \second{0.334} & 0.538  \\
MeanFlow~\cite{geng2025mean} 
  & 0.498 & 20.85 & 3.086 & \second{0.249} & 0.197& 0.292 & 0.563  \\
\midrule
\rowcolor{orange}
LangRetrieval (Ours)
  & \best{0.594} & \best{22.66} & \best{2.555} & \best{0.206} &\best{0.268} & \best{0.383} & \best{0.379} \\
\rowcolor{gray!20}
\textcolor{red}{\textit{Improv. vs 2\textsuperscript{nd}}}
  & \textcolor{red}{+0.7\%} & \textcolor{red}{+5.6\%} & \textcolor{red}{+1.6\%} & \textcolor{red}{+14.9\%} & \textcolor{red}{+17.0\%} & \textcolor{red}{+14.7\%} & \textcolor{red}{+0.3\%} \\
\rowcolor{gray!20}
\textcolor{red}{\textit{Improv. vs MeanFlow}}
  & \textcolor{red}{+16.5\%} & \textcolor{red}{+7.5\%} & \textcolor{red}{+14.5\%} & \textcolor{red}{+14.9\%} & \textcolor{red}{+36.0\%} & \textcolor{red}{+31.2\%} & \textcolor{red}{+32.7\%} \\
\bottomrule
\end{tabular}}
\end{table}

\begin{table*}[t]
\centering
\caption{Efficiency comparison of all methods.
Params denotes the total number of trainable parameters.
FLOPs denotes the total floating-point operations for a single inference.
Time is measured on a single GPU with batch size 1, averaged over 100 runs.
$^\dagger$Params and FLOPs of LangRetrieval exclude the frozen CLIP ViT-L/14 encoder (427.62M / 61.60G).
\textbf{Bold}: best; \second{underline}: second best.
}
\label{tab:efficiency}
\renewcommand{\arraystretch}{1.15}
\setlength{\tabcolsep}{5pt}
\begin{tabular}{ccccc|ccccc}
\toprule
\textbf{Method} & \textbf{Steps} & \textbf{FLOPs (G)} & \textbf{Time (ms)} & \textbf{Params (M)} & \textbf{LPIPS}$\downarrow$& \textbf{PSNR}$\uparrow$& \textbf{SSIM}$\uparrow$& \textbf{CSI@30}$\uparrow$& \textbf{HSS@30}$\uparrow$\\
\midrule
AA-TransUNet~\cite{yang2022aa}
& 1&3.78  &6.7&39.88 
& 0.280&20.49&0.539&0.130&0.223\\
EarthFormer~\cite{gao2022earthformer} 
&    1&     2.53  &    13.4&    8.68  &0.338&20.86&0.555&0.145&0.245\\
Smaat-UNet~\cite{trebing2021smaat}          
&    1&     2.40  &     3.6 &    4.02 &0.333&19.18&0.501&0.118&0.199\\
DiffCast~\cite{yu2024diffcast}            
&  250&  7459.11  &3302.8&46.25  
&0.274&21.23&0.590&0.174&0.289\\
Pix2Pix~\cite{akter2024generative}             
&    1&     8.25  &     7.4&   31.36  &0.257&19.88&0.523&0.189&0.305\\
MeanFlow~\cite{geng2025mean}             
&1&12.96  &22.3&  107.80  
&0.249&20.85&0.498&0.148&0.249\\
\midrule
\rowcolor{orange}
\textbf{LangRetrieval (Ours)}$^\dagger$ & 1& 33.87& 27.3& 45.21  
&\best{0.206}&\best{22.66}&\second{0.594}&\best{0.228}&\best{0.349}\\
\rowcolor{orange}
\textbf{LangRetrieval (Ours)}$^\dagger$ & 5& 169.39& 97.6 & 45.21 
&\second{0.208}&\second{22.57}&\best{0.599}&\second{0.221}&\second{0.341}\\
\rowcolor{orange}
\textbf{LangRetrieval (Ours)}$^\dagger$ & 10&338.71& 186.5&45.21 
&0.210&22.48&0.586&0.219&0.338\\
\rowcolor{orange}
\textbf{LangRetrieval (Ours)}$^\dagger$  & 20& 677.18 & 354.7& 45.21  &0.212&22.41&0.580&0.218&0.337\\
\bottomrule
\end{tabular}
\end{table*}

\begin{figure*}[ht]
    \centering
    \includegraphics[width=0.98\textwidth]{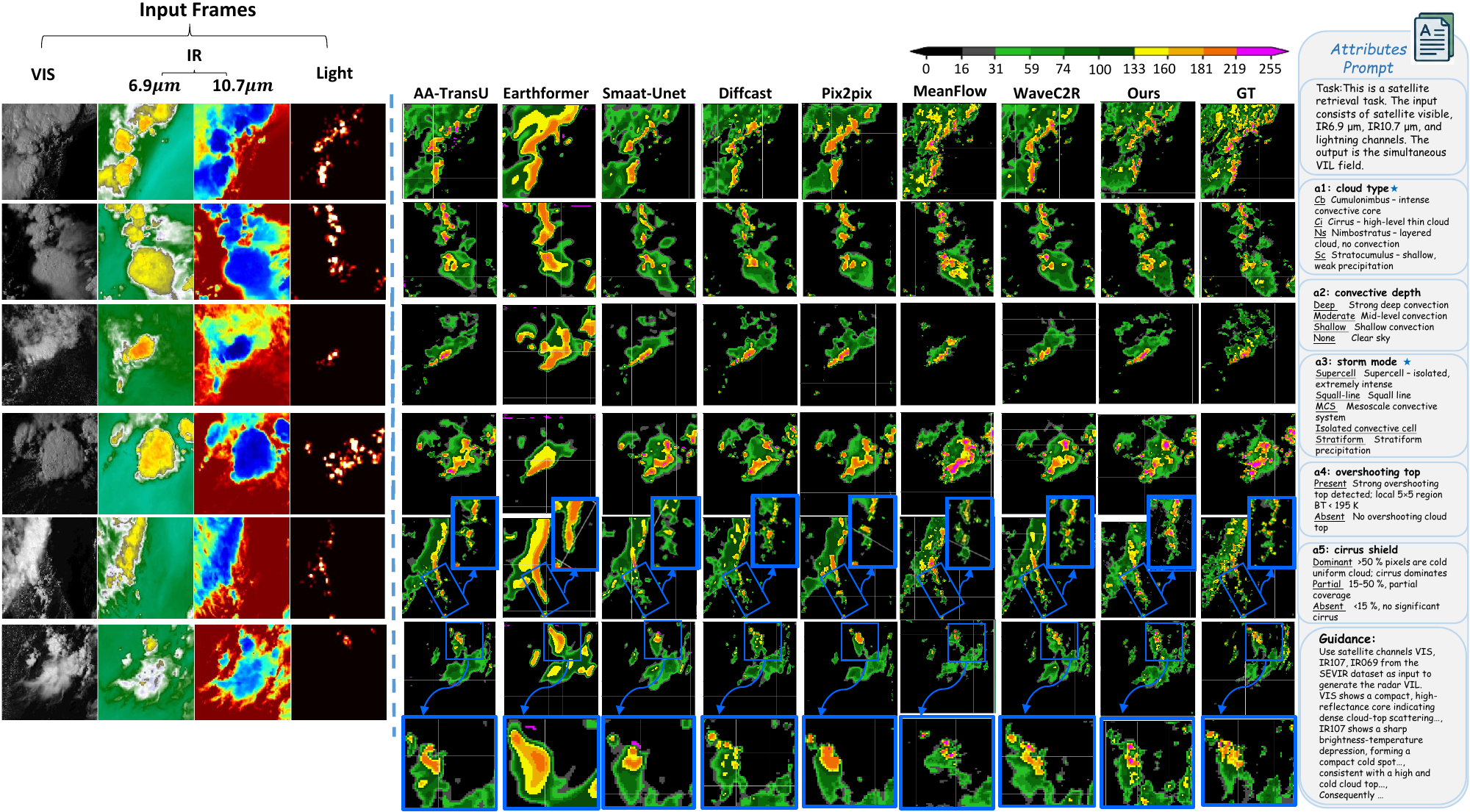}
    \caption{
        Qualitative comparison of radar reflectivity predictions for six heavy precipitation events from the SEVIR dataset. Red boxes highlight magnified convective regions where our LangRetrieval demonstrates superior intensity accuracy. Input modalities include visible, dual-channel infrared, and lightning observations.
    }
    \label{fig:sevir_vis}
\end{figure*}
\begin{figure*}[ht]
    \centering
    \includegraphics[width=\textwidth]{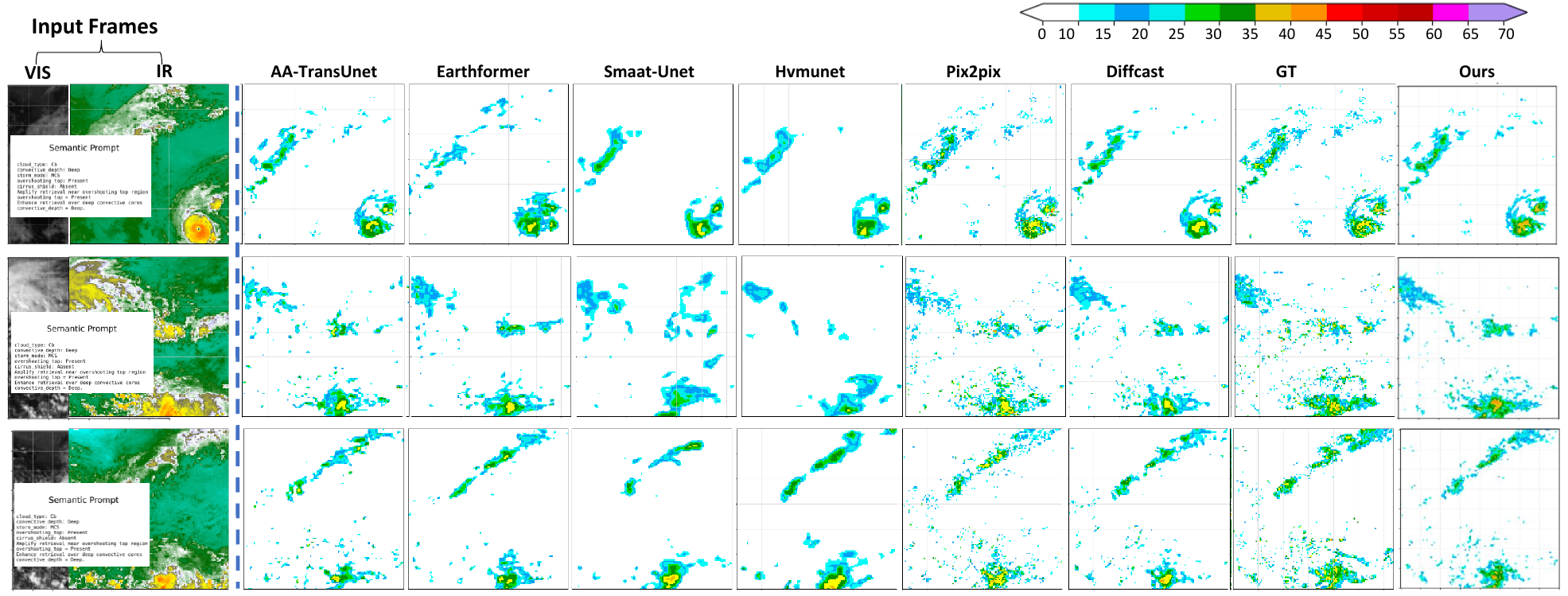}
    \caption{
        Qualitative comparison of satellite-to-radar retrieval results on the Southeast China FY-4B test set.
        From top to bottom: input satellite imagery (VIS and IR channels), outputs of AA-TransUnet,
        Earthformer, Smaat-Unet, Hvmunet, Pix2Pix, DiffCast, Ground Truth (GT), and our proposed
        LangRetrieval. The semantic prompt describes key meteorological features (\emph{e.g.},
        deep convective cores), guiding our model to generate more accurate radar reflectivity fields.
        Color bar (dBZ) ranges from 0 to 70.
    }
    \label{fig:fy4b_vis}
  \vspace{-5pt}
\end{figure*}

\noindent\textbf{Evaluation Metrics.} 
Following established practices in weather nowcasting~\cite{shi2026wavec2r}, we employ multiple evaluation metrics. For meteorological assessment, Critical Success Index (CSI) and Heidke Skill Score (HSS) are computed at multiple thresholds. For SEVIR, thresholds are \{74, 133, 160, 181, 219\} kg/m$^2$; for FY-4B, thresholds are \{10, 20, 25, 30, 35\} dBZ. We report mean CSI (Avg.CSI) and mean HSS (Avg.HSS) for overall accuracy. For image quality, we report Structural Similarity Index Measure (SSIM), Peak Signal-to-Noise Ratio (PSNR), Mean Absolute Error (MAE), and Learned Perceptual Image Patch Similarity (LPIPS). Additionally, False Alarm Rate (FAR) is reported for FY-4B.

\noindent\textbf{Baseline Methods.} 
We compare LangRetrieval against six competitive baselines: AA-TransUnet (CNN-Transformer hybrid), Earthformer (meteorology-oriented), Smaat-Unet (CNN-based), Diffcast (diffusion-based generative), Pix2Pix (GAN-based generative), and MeanFlow (flow-based).

\noindent\textbf{Implementation Details.} 
Our framework LangRetrieval is trained in two stages on four NVIDIA GeForce RTX 3090 GPUs.
In Stage~I, $U_\theta$ network is optimized with AdamW ($\text{lr}=3\times10^{-4}$, weight decay $10^{-5}$) for 80K iterations with batch size 16.
In Stage~II, the attribute policy $\pi_\phi$ is first warm-started via supervised cross-entropy on VLM-annotated samples for 5K steps ($\text{lr}=10^{-3}$), then refined by GRPO over $T\!=\!3$ rounds of 10K steps each.
Both $\pi_\phi$ and $U_\theta$ are jointly optimized during Phase 2b: the policy uses $\text{lr}\!=\!10^{-4}$ (group size $K\!=\!8$, KL coefficient $\beta\!=\!0.04$), sampling temperature $T_s\!=\!0.8$.
All inputs are resized to $128\!\times\!128$. Further architecture and training details are provided in the supplementary material.

\subsection{Main Results}

\noindent\textbf{Quantitative Performance.}
Table~\ref{tab:sota} presents quantitative comparisons on 
the SEVIR dataset. LangRetrieval achieves an average CSI 
of 0.360, outperforming the second-best method (Diffcast, 
0.310) by 16.1\%. Improvements are most pronounced at high 
precipitation thresholds: CSI@219 reaches 0.185 (+46.8\% 
vs.\ Diffcast) and CSI@133 reaches 0.405 (+24.7\% vs.\ 
the second-best), demonstrating that scene-level semantic 
guidance effectively resolves the many-to-one ambiguity 
between visually similar cloud-top signatures and distinct 
surface precipitation regimes that purely pixel-driven 
methods fail to disambiguate.
Avg.HSS reaches 0.467 (+4.2\% vs.\ Diffcast), confirming 
that the gains are not obtained at the cost of inflated 
false alarms. LPIPS is 0.202 (best), improving by 14.4\% 
relative to Diffcast (0.236), suggesting that 
language-guided generation yields more physically consistent 
precipitation structures alongside improved detection skill.

Table~\ref{tab:sota_fy} demonstrates consistent 
improvements on the geographically distinct Southeast 
China FY-4B dataset. LangRetrieval attains strong 
operational gains across thresholds: CSI@20 is 0.309 
(+15.7\% vs.\ the second-best) and CSI@25 is 0.282 
(+24.2\% vs.\ the second-best). Avg.CSI on FY-4B is 
0.268 and Avg.HSS is 0.383, representing substantial 
improvements over competitive baselines (Avg.CSI +17.0\% 
and Avg.HSS +14.7\% vs.\ the second-best). FAR analysis 
indicates a favorable detection-to-false-alarm trade-off: 
FAR@20 is 0.303 (better than Diffcast's 0.316, i.e., 
$\sim$4.1\% improvement), and Avg.FAR is 0.379 (best). 
These results demonstrate robust cross-domain 
generalization across different spatial resolutions 
(1~km for SEVIR vs.\ 4~km for FY-4B) and climatologies.

\noindent\textbf{Perceptual Quality and Metric Balance.}
Table~\ref{tab:sota_perceptual} shows that LangRetrieval also leads in image-quality metrics on FY-4B: SSIM = 0.594, PSNR = 22.66, MAE = 2.555, and LPIPS = 0.206 (all best). Compared with Diffcast, PSNR improves by 5.6\% and SSIM is slightly higher (+0.7\%). LPIPS is substantially improved relative to flow- and baseline methods (e.g., +14.9\% vs. MeanFlow). Crucially, LangRetrieval simultaneously improves meteorological skill (Avg.CSI, Avg.HSS) while maintaining or improving perceptual fidelity, indicating the GRPO-driven semantic optimization refines task-relevant structures without degrading visual realism.

\noindent\textbf{Computational Efficiency and Deployment 
Feasibility.}
Table~\ref{tab:efficiency} highlights computational 
efficiency. For a full 20-step ODE integration, 
LangRetrieval requires 354.7~ms, yielding a $9.3\times$ 
speedup over Diffcast (3302.8~ms). LangRetrieval supports 
flexible inference budgets: 1-step inference is 27.3~ms, 
5-step is 97.6~ms, 10-step is 186.5~ms, and 20-step is 
354.7~ms, enabling dynamic latency--accuracy trade-offs 
without architectural changes.
Crucially, at test time the semantic tokens are generated 
by the lightweight policy $\pi_\phi$ in a single forward 
pass without any VLM call, making the full pipeline 
deployable on resource-constrained operational platforms 
where large foundation model inference is unavailable. 
With 45.21M trainable parameters (excluding the frozen 
CLIP encoder), the model remains parameter-efficient 
among the baselines.

\noindent\textbf{Qualitative Analysis and Language-Guided Precipitation Generation.}
Fig.~\ref{fig:sevir_vis} and Fig.~\ref{fig:fy4b_vis} provide visual evidence of the quantitative improvements. 
The language guidance encodes five meteorologically-meaningful satellite-derived attributes: 
\textit{cloud type} (Cb: cumulonimbus for intense cores, Ci: cirrus for high clouds, Ns: nimbostratus for sustained rain, Sc: stratocumulus for weak precipitation); 
\textit{convective depth} (Deep/Moderate/Shallow/None); 
\textit{storm mode} (Supercell, Squall-line, MCS, Isolated cell, Stratiform rain); 
\textit{overshooting top} (presence of tropopause-penetrating thunderstorms, BT$<$195K in local 5$\times$5 region); and 
\textit{cirrus shield} (dominant >50\%, partial 15--50\%, or absent <15\% coverage-a primary source of false alarms).

Three key qualitative advantages emerge across both datasets. 
\textit{First}, LangRetrieval accurately localizes extreme precipitation cores (high threshold), whereas deterministic methods (AA-TransUnet, Earthformer, Smaat-Unet) produce overly smooth predictions with underestimated intensity, and generative methods (Diffcast, Pix2Pix) introduce spurious scattered precipitation in clear-sky regions. 
\textit{Second}, the language constraints-particularly cloud type and cirrus shield attributes-effectively suppress false alarms in meteorologically implausible regions, a problem that unconstrained generative models struggle with. 
\textit{Third}, LangRetrieval better preserves vertical convective structure as inferred from multi-spectral satellite channels (VIS, IR 6.9\,$\mu$m, IR 10.7\,$\mu$m), as evidenced by the VIL maps which show coherent organization matching ground truth, while baseline predictions lack spatial consistency (Fig.~\ref{fig:sevir_vis}).

On FY-4B (Fig.~\ref{fig:fy4b_vis}), these patterns persist: LangRetrieval generates sharper convective boundaries and more realistic mesoscale precipitation organization compared to all baselines. 
This consistency across datasets with different resolutions and geographic domains demonstrates that language-guided conditioning provides a robust inductive bias for S2R.

\begin{figure*}[t]
  \centering
  \includegraphics[width=1\linewidth]{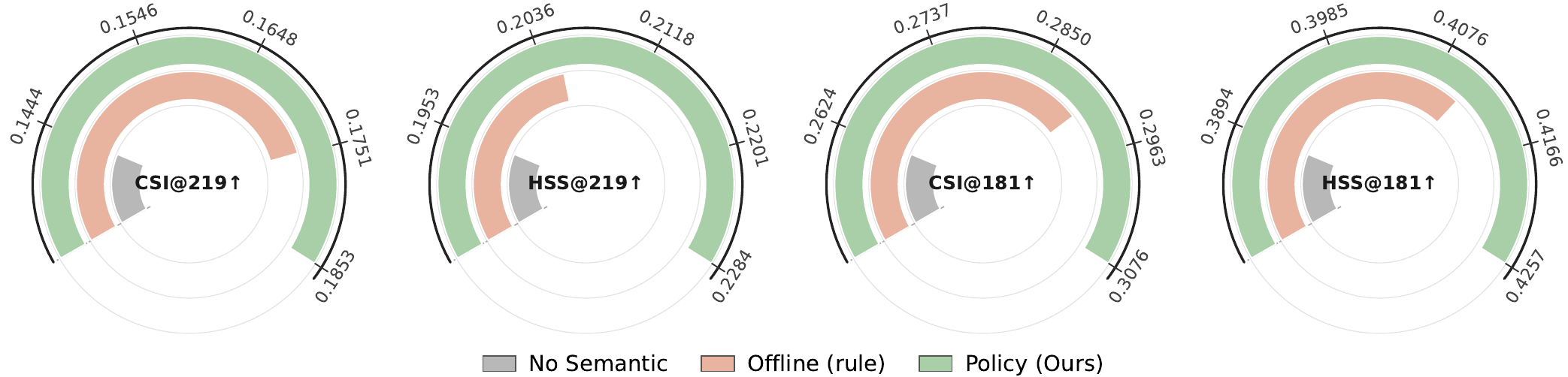}
  \caption{Component-level ablation of semantic injection modes on SEVIR. 
  We compare three strategies: 
  (i)~\textit{No Semantic}, an unconditioned baseline with empty text prompt; 
  (ii)~\textit{Offline (rule-based)}, attributes derived from fixed 
  meteorological thresholds; and 
  (iii)~\textit{Policy (Ours)}, attributes learned via policy optimization 
  ($\pi_\phi$ after IRR-3). The learned policy achieves the best performance 
  across high- and mid-threshold precipitation skill metrics.}
  \label{fig:policy}
  \vspace{-10pt}
\end{figure*}

\subsection{Ablation Study}

\noindent\textbf{Effect of Semantic Conditioning Strategy.}
To assess the contribution of policy-based attribute prediction, 
we compare three conditioning variants on the public SEVIR dataset (see Fig.~\ref{fig:policy}),
holding all other components fixed.

The unconditioned baseline (\textit{No Semantic}) receives no 
language guidance and yields CSI@219\,=\,0.1052 and HSS@219\,=\,0.1457.
Replacing this with rule-derived attributes (\textit{Offline}) improves 
detection skill: CSI@219\,=\,0.1253 ($+$19.1\%) and HSS@219\,=\,0.1824 ($+$25.2\%), 
confirming that semantic structure carries useful meteorological information.

Our \textit{Policy} variant learns the attribute mapping end-to-end 
under the CSI reward, achieving CSI@219\,=\,0.1471 ($+$17.3\% over Offline) 
and HSS@219\,=\,0.2278 ($+$24.8\%).
Consistent gains across both metrics demonstrate that learned attributes 
are meteorologically discriminative, whereas rigid threshold rules cannot 
adapt to convective morphology diversity.

\begin{figure}[t]
  \centering
  \includegraphics[width=1\linewidth]{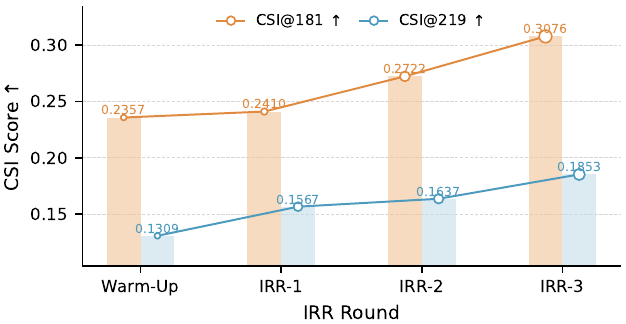}
\caption{Ablation studies: Iterative Refinement Rounds (IRR) on SEVIR.
Progressive application of IRR substantially improves high-threshold detection performance compared to Warm-Up.}
  \label{fig:irr_dynamics}
  \vspace{-15pt}
\end{figure}

\noindent\textbf{Iterative Refinement Round (IRR) Dynamics.}
Fig.~\ref{fig:irr_dynamics} validates CSI-driven self-evolution across three 
GRPO rounds on SEVIR. After Phase~2a warm-up, the policy achieves baseline 
CSI@181\,=\,0.2357 and CSI@219\,=\,0.1309.
Each IRR advances both metrics monotonically: IRR-3 converges to CSI@181\,=\,0.3076 ($+$30.5\%) and CSI@219\,=\,0.1853 ($+$41.5\%).
Notably, extreme-precipitation gains (CSI@219) dominate in IRR-1 ($+$19.7\%), indicating rapid discovery of severe-weather-critical attributes. 
The consistent multi-threshold improvement demonstrates that self-evolved meteorological attributes, optimized under retrieval-aware rewards, systematically enhance detection across precipitation intensity regimes 
and substantially surpass the VLM teacher initialized in Phase~2a.


\section{Conclusion}
This work introduces a paradigm for satellite-to-radar precipitation retrieval that treats \textit{meteorological semantics as a learnable, task-optimized variable} rather than a fixed auxiliary input. The core innovation lies in establishing closed-loop optimization between two 
historically disconnected pathways: semantic generation and retrieval accuracy. 
We demonstrate that the semantic gap inherent in precipitation mapping can be systematically bridged through two coupled mechanisms. First, semantic warm-up (Stage~I) anchors learned 
attributes to vision-language knowledge via attention-guided conditioning, providing fine-grained semantic-visual grounding. 
Second, self-evolving semantic optimization (Stage~II) replaces expensive annotation-driven semantics with a lightweight learnable policy that directly optimizes under retrieval-aware rewards. Extensive validation on FY-4B and SEVIR demonstrates that LangRetrieval achieves state-of-the-art performance, with consistent multi-threshold improvements confirming that self-evolved semantics, optimized under task rewards, systematically enhance retrieval across precipitation intensity regimes.

\clearpage
\bibliographystyle{IEEEtran}
\bibliography{egbib}

\clearpage
\setcounter{page}{1}
\section*{Appendix}

This supplementary material is organised as follows.
Appendix~A formalises the deterministic rules used to compute
the three pixel-verifiable attributes ($a_2, a_4, a_5$) from
the IR brightness temperature field.
Appendix~B provides the complete training configurations for
Stage~I (Semantic Warm-Up) and Stage~II (GRPO Policy
Optimisation), including all hyperparameters, reward weights,
and hardware specifications.
Appendix~C presents the multi-VLM annotation fusion ablation
(V3), which investigates whether combining annotations from
Qwen-VL-Max, Kimi-VL, and DeepSeek-VL via the proposed
TASRE framework improves upon the single-source configuration
adopted in the main experiment.
Appendix~D discusses the practical significance of
satellite-based radar retrieval in observationally-sparse
regions, with cross-regional transfer results on Xinjiang
Province and oceanic coverage analysis for southeastern China.

\subsection{Pixel-verifiable attributes ($a_2, a_4, a_5$).}
\label{app:pixel_rules}
Let $\mathrm{BT}$ denote the brightness temperature field derived from
the normalised IR channel.

\vspace{3pt}\noindent
\emph{Convective depth} is determined by the spatial gradient statistics of BT:
\begin{equation}
  a_2 = \mathcal{C}\!\left(
    \mathrm{P}_{95}\!\bigl(\lVert\nabla\, \mathrm{BT}\rVert\bigr)
  \right),
  \label{eq:conv_depth}
\end{equation}
where $\mathcal{C}$ is a threshold classifier mapping gradient magnitude to
four ordinal categories.

\vspace{3pt}\noindent
\emph{Overshooting top} is detected via localised cold-point anomalies:
\begin{equation}
  a_4 =
  \begin{cases}
    \text{Present} & \text{if } \min_{\Omega}(\mathrm{BT}) < \tau_{\rm ot}, \\
    \text{Absent}  & \text{otherwise,}
  \end{cases}
  \label{eq:ot}
\end{equation}
where $\Omega$ denotes a local neighbourhood and $\tau_{\rm ot}$ is a
temperature threshold.

\vspace{3pt}\noindent
\emph{Cirrus shield} is identified by low-texture cold regions:
\begin{equation}
  a_5 = \mathcal{C}'\!\left(
    \frac{|\{(i,j): \sigma^2_\Omega(i,j) < \epsilon \;\wedge\;
          \mathrm{BT}(i,j) < \tau_{\rm ci}\}|}{HW}
  \right),
  \label{eq:cirrus}
\end{equation}
where $\sigma^2_\Omega$ is the local variance and $\mathcal{C}'$ maps coverage
fraction to three ordinal categories.
\subsection{Training Details}
\subsubsection{Stage I: Semantic Warm-Up Training Details}
\label{app:stage1}

Stage~I trains the conditional flow matching retrieval backbone $U_\theta$ with
semantic cross-attention, using offline rule-based prompts. The three
pixel-verifiable attributes (convective depth, overshooting top, cirrus shield)
are computed directly from the satellite imagery, while the two VLM-dependent
attributes (cloud type, storm mode) are set to sentinel values.

Table~\ref{tab:stage1_config} summarises the Stage~I hyperparameters.
\begin{table}[h]
\centering
\caption{Stage~I (Semantic Warm-Up) training configuration.}
\label{tab:stage1_config}
\resizebox{0.5\textwidth}{!}{
\vspace{4pt}
\small
\begin{tabular}{ll}
\toprule
\textbf{Hyperparameter} & \textbf{Value} \\
\midrule
\multicolumn{2}{l}{\textit{Data}} \\
\quad Dataset & FY-4B (Southeast China) \\
\quad Input channels & 3 (VIS, IR069, IR107) \\
\quad Output & Radar reflectivity, 1 channel \\
\quad Resolution & $128 \times 128$ \\
\quad Normalisation & $[0, 1]$; physical value $= \text{pixel} \times 70$~dBZ \\
\midrule
\multicolumn{2}{l}{\textit{Model Architecture}} \\
\quad Backbone & VelocityUNet \\
\quad UNet base channels & 64 \\
\quad Channel multipliers & $[1, 2, 4, 8]$ \\
\quad Cross-attention resolutions & $\{32, 64\}$ (5 injection points) \\
\quad Cross-attention $d_k$ & 64 \\
\quad CLIP encoder & ViT-L/14 (frozen), $d = 768$ \\
\quad Total parameters & $\sim$44.8M (incl.\ cross-attention $\sim$1.3M) \\
\midrule
\multicolumn{2}{l}{\textit{Flow Matching}} \\
\quad Time distribution & Log-normal ($\mu = -0.4,\; \sigma = 1.0$) \\
\quad Flow ratio & 0.50 \\
\quad NFE (inference) & 20 (Euler ODE) \\
\midrule
\multicolumn{2}{l}{\textit{Training}} \\
\quad Loss & $\mathcal{L}_{\text{CFM}}$ \\
\quad Optimiser & AdamW (weight decay $10^{-5}$) \\
\quad Learning rate & $3 \times 10^{-4}$ \\
\quad Batch size & 16 \\
\quad Training steps & 80{,}000 \\
\quad VLM loss weight $\lambda$ & 0.1 \\
\quad Hardware & $4 \times$ NVIDIA GeForce RTX 3090 (24\,GB) \\
\bottomrule
\end{tabular}}
\end{table}

\subsubsection{Stage II: Policy Optimisation Training Details}
\label{app:stage2}

Stage~II trains a lightweight policy network $\pi_\phi$ (AttributePredictor,
$\sim$1.7M parameters) that maps satellite images to five-attribute semantic
prompts. It consists of two phases.

\paragraph{Phase~IIa: Supervised Pre-Training.}
$\pi_\phi$ is initialised via cross-entropy loss on VLM annotations
(Eq.~\ref{eq:pretrain_loss} in main text). In the main experiment, we use
\textbf{single-source Qwen-VL-Max} annotations exclusively.

\paragraph{Phase~IIb: GRPO Iterative Refinement.}
With the Stage~I backbone frozen, $\pi_\phi$ is optimised with Group Relative
Policy Optimisation (GRPO) across $N = 3$ rounds. At the end of each round, the
reference policy is progressively reset.

Table~\ref{tab:stage2_config} presents the complete Stage~II configuration.

\begin{table}[h]
\centering
\caption{Stage~II (GRPO Policy Optimisation) training configuration --- main experiment.}
\label{tab:stage2_config}
\resizebox{0.5\textwidth}{!}{
\vspace{4pt}
\small
\begin{tabular}{ll}
\toprule
\textbf{Hyperparameter} & \textbf{Value} \\
\midrule
\multicolumn{2}{l}{\textit{Policy Network ($\pi_\phi$)}} \\
\quad Architecture & CNN encoder + GAP + per-attribute linear heads \\
\quad Input channels & 3 (same as satellite) \\
\quad Parameters & $\sim$1.7M \\
\quad Number of attributes & 5 \\
\midrule
\multicolumn{2}{l}{\textit{Phase IIa: Supervised Pre-Training}} \\
\quad Annotation source & Qwen-VL-Max (single VLM) \\
\quad Loss & Cross-entropy on 5-attribute labels \\
\quad Steps & 5{,}000 \\
\quad Learning rate & $1 \times 10^{-3}$ \\
\quad Optimiser & Adam \\
\quad Gradient clipping & 1.0 \\
\midrule
\multicolumn{2}{l}{\textit{Phase IIb: GRPO Iterative Refinement}} \\
\quad IRR rounds ($N$) & 3 \\
\quad Steps per round & 10{,}000 \\
\quad Group size ($K$) & 8 \\
\quad Sampling temperature ($T_s$) & 0.8 \\
\quad KL penalty ($\beta$) & 0.04 \\
\quad Learning rate & $1 \times 10^{-4}$ \\
\quad Optimiser & Adam \\
\quad Gradient clipping & 1.0 \\
\quad Reference reset & Progressive (deepcopy after each round) \\
\midrule
\multicolumn{2}{l}{\textit{Common}} \\
\quad Batch size & 8 \\
\quad NFE (ODE steps) & 20 \\
\quad Backbone $U_\theta$ & Frozen (Stage~I checkpoint) \\
\quad CLIP encoder & Frozen (ViT-L/14) \\
\midrule
\multicolumn{2}{l}{\textit{CSI Reward Weights ($w_\tau$)}} \\
\quad $\tau = 10$~dBZ & 0.15 \\
\quad $\tau = 20$~dBZ & 0.15 \\
\quad $\tau = 25$~dBZ & 0.20 \\
\quad $\tau = 30$~dBZ & 0.25 \\
\quad $\tau = 35$~dBZ & 0.15 \\
\quad $\tau = 40$~dBZ & 0.10 \\
\bottomrule
\end{tabular}}
\end{table}

\paragraph{Evaluation Protocol.}
All models are evaluated on the held-out test set (1{,}114 samples) with
NFE $= 20$ and batch size 16. We report CSI, FAR, POD, and HSS at thresholds
$\tau \in \{10, 20, 25, 30, 35, 40\}$~dBZ, along with pixel (PSNR, MAE) and
perceptual (SSIM, LPIPS) metrics.

\subsection{Multi-VLM Annotation Fusion Ablation (V3)}
\label{app:multi_vlm}

A natural question is whether combining annotations from multiple VLMs can
further improve the quality of pseudo-labels used in Phase~IIa, thereby
producing a better-initialised policy for GRPO. We investigate this through the
\textbf{TASRE} (Task-Aligned Source Reliability Evaluation) framework, which
evaluates and fuses annotations from three state-of-the-art VLMs using the
downstream CSI metric as the reliability signal.

\subsubsection{TASRE Framework}

Given $M$ VLM annotation sources $\{s_1, \ldots, s_M\}$, TASRE assigns fusion
weights proportional to each source's downstream retrieval performance:
\begin{equation}
  w_m = \frac{\exp(\mathrm{CSI}_m / T)}{\sum_{j=1}^M \exp(\mathrm{CSI}_j / T)},
  \label{eq:tasre}
\end{equation}
where $\mathrm{CSI}_m$ is the mean CSI score achieved when using source $s_m$'s
annotations exclusively with the frozen Stage~I backbone, and $T$ is a softmax
temperature controlling the sharpness of the weight distribution.

We evaluate three VLMs:
\begin{itemize}[nosep,leftmargin=1.5em]
  \item \textbf{Qwen-VL-Max}: Alibaba's multimodal model (used in the main experiment)
  \item \textbf{Kimi-VL}: Moonshot AI's vision-language model
  \item \textbf{DeepSeek-VL}: DeepSeek's multimodal model
\end{itemize}

In addition to TASRE, we compare four alternative fusion strategies:
\begin{table}[h]
\centering
\caption{Five annotation fusion strategies evaluated in the V3 ablation.}
\label{tab:fusion_strategies}
\resizebox{0.5\textwidth}{!}{
\vspace{4pt}
\small
\begin{tabular}{lp{9cm}}
\toprule
\textbf{Strategy} & \textbf{Description} \\
\midrule
\textbf{Single} & Use only Qwen-VL-Max annotations (equivalent to V2 main experiment) \\
\textbf{Majority Vote} & Per-attribute majority voting across three VLMs \\
\textbf{Uniform} & Equal-weight ($\frac{1}{3}$) averaging of VLM confidence scores \\
\textbf{Confidence} & Self-reported confidence-weighted fusion \\
\textbf{TASRE} & Task-aligned CSI-weighted fusion (Eq.~\ref{eq:tasre}) \\
\bottomrule
\end{tabular}}
\end{table}

\subsubsection{VLM Source Reliability Evaluation}

Table~\ref{tab:vlm_csi} reports the per-source CSI scores and the resulting
TASRE fusion weights under three temperature settings. All evaluations use the
frozen Stage~I backbone on the validation set.
\begin{table}[h]
\centering
\caption{VLM source reliability measured by downstream CSI and corresponding
  TASRE weights under different temperatures. All three VLMs achieve comparable
  performance ($\sim$0.242 CSI), resulting in nearly uniform weights regardless
  of temperature.}
\label{tab:vlm_csi}
\resizebox{0.5\textwidth}{!}{
\vspace{4pt}
\small
\begin{tabular}{lccc|ccc}
\toprule
\multirow{2}{*}{\textbf{VLM Source}} & \multicolumn{3}{c|}{\textbf{CSI Score}} & \multicolumn{3}{c}{\textbf{TASRE Weight}} \\
\cmidrule(lr){2-4} \cmidrule(lr){5-7}
& $T{=}0.5$ & $T{=}1.0$ & $T{=}2.0$ & $T{=}0.5$ & $T{=}1.0$ & $T{=}2.0$ \\
\midrule
Qwen-VL-Max     & 0.2427 & 0.2420 & 0.2423 & 0.3334 & 0.3332 & 0.3334 \\
Kimi-VL         & 0.2424 & 0.2428 & 0.2423 & 0.3332 & 0.3334 & 0.3334 \\
DeepSeek-VL     & 0.2428 & 0.2427 & 0.2420 & 0.3334 & 0.3334 & 0.3330 \\
\midrule
\textbf{Std Dev} & 0.0002 & 0.0004 & 0.0001 & 0.0001 & 0.0001 & 0.0002 \\
\bottomrule
\end{tabular}}
\end{table}

\textbf{Key observation:} All three VLMs achieve nearly identical CSI scores
($0.242 \pm 0.0004$), yielding approximately uniform TASRE weights
($\sim$0.333) across all temperatures. This indicates that, for this particular
five-attribute annotation task on satellite imagery, the three VLMs produce
annotations of equivalent quality when measured by the downstream retrieval
metric.

\subsubsection{V3 Ablation Training Configuration}

To efficiently compare fusion strategies, we use a faster configuration for the
ablation runs while keeping the Stage~I backbone identical.
\begin{table}[h]
\centering
\caption{Training configuration comparison: main experiment (V2) vs.\ V3 ablation.
  Red values highlight differences.}
\label{tab:v2_v3_config}
\resizebox{0.5\textwidth}{!}{
\vspace{4pt}
\small
\begin{tabular}{lcc}
\toprule
\textbf{Hyperparameter} & \textbf{V2 (Main Experiment)} & \textbf{V3 (Ablation)} \\
\midrule
\multicolumn{3}{l}{\textit{Phase IIa: Supervised Pre-Training}} \\
\quad Annotation source & Qwen only & 5 strategies (Table~\ref{tab:fusion_strategies}) \\
\quad Pretrain steps & 5{,}000 & 5{,}000 \\
\quad Learning rate & $1 \times 10^{-3}$ & $1 \times 10^{-3}$ \\
\midrule
\multicolumn{3}{l}{\textit{Phase IIb: GRPO}} \\
\quad IRR rounds ($N$) & 3 & 3 \\
\quad Steps per round & 10{,}000 & \textcolor{red}{500} \\
\quad Group size ($K$) & 8 & \textcolor{red}{4} \\
\quad Sampling temperature ($T_s$) & 0.8 & \textcolor{red}{0.1} \\
\quad KL penalty ($\beta$) & 0.04 & \textcolor{red}{0.01} \\
\midrule
\multicolumn{3}{l}{\textit{Common}} \\
\quad Batch size & 8 & 8 \\
\quad NFE (ODE steps) & 20 & 20 \\
\quad Stage~I backbone & Frozen & Frozen \\
\quad TASRE temperature & --- & $T \in \{0.5, 1.0, 2.0\}$ \\
\bottomrule
\end{tabular}}
\end{table}

\textbf{Rationale:} The V3 ablation uses fewer steps per round (500 vs.\ 10{,}000)
and more aggressive sampling ($T_s = 0.1$) to rapidly identify \emph{relative}
performance trends across fusion strategies, rather than achieving the absolute
optimal performance obtained in the main experiment.

\subsubsection{Experimental Design}

The V3 ablation consists of two experiment groups:

\textbf{Experiment~1: Fusion Strategy Comparison (Fixed $T = 1.0$).}
We train five independent runs, each using a different fusion strategy from
Table~\ref{tab:fusion_strategies}, all with TASRE temperature $T = 1.0$.
This isolates the effect of annotation source / fusion method from temperature.

\textbf{Experiment~2: TASRE Temperature Ablation.}
For the TASRE strategy specifically, we vary the softmax temperature
$T \in \{0.5, 1.0, 2.0\}$:
\begin{itemize}[nosep,leftmargin=1.5em]
  \item $T = 0.5$: Sharp distribution, biasing toward the highest-CSI source
  \item $T = 1.0$: Standard softmax (default)
  \item $T = 2.0$: Smooth distribution, approaching uniform weighting
\end{itemize}

All models are evaluated identically: CSI, FAR, POD, HSS at
$\tau \in \{10, 20, 25, 30, 35, 40\}$~dBZ on the test set (1{,}114 samples),
with NFE $= 20$ and evaluation batch size 16.

\subsubsection{Results and Discussion}

\paragraph{Experiment 1: Fusion Strategy Comparison.}

Table~\ref{tab:fusion_results} reports the retrieval performance of all five
fusion strategies under identical GRPO training (with fast ablation settings).
Mean values are computed over thresholds $\tau \in \{10, 20, 25, 30, 35, 40\}$~dBZ.
\begin{table}[h]
\centering
\caption{Fusion strategy comparison (fixed $T = 1.0$).
  Mean metrics are averaged over all six thresholds.
  \textbf{Bold}: best per column.
  Bottom row: range across strategies.}
\label{tab:fusion_results}
\resizebox{0.5\textwidth}{!}{
\vspace{4pt}
\small
\setlength{\tabcolsep}{4pt}
\begin{tabular}{l cccc}
\toprule
\textbf{Strategy}
& \textbf{CSI-Mean}$\uparrow$
& \textbf{FAR-Mean}$\downarrow$
& \textbf{POD-Mean}$\uparrow$
& \textbf{HSS-Mean}$\uparrow$ \\
\midrule
Single (Qwen)
& 0.2238 & 0.4756 & 0.2719 & 0.3254 \\
Majority Vote
& 0.2241 & 0.4697 & 0.2722 & 0.3262 \\
Uniform
& 0.2240 & 0.4700 & 0.2721 & 0.3259 \\
Confidence
& 0.2234 & 0.4726 & 0.2713 & 0.3249 \\
TASRE ($T{=}1.0$)
& 0.2239 & 0.4701 & 0.2720 & 0.3259 \\
\midrule
\textit{Max $-$ Min}
& \textit{0.0007} & \textit{0.0059} & \textit{0.0009} & \textit{0.0013} \\
\bottomrule
\end{tabular}}
\end{table}

Table~\ref{tab:fusion_per_threshold} provides the per-threshold CSI breakdown.
\begin{table}[h]
\centering
\caption{Per-threshold CSI comparison across fusion strategies (fixed $T = 1.0$).
  \textbf{Bold}: best per column.}
\label{tab:fusion_per_threshold}
\resizebox{0.5\textwidth}{!}{
\vspace{4pt}
\small
\setlength{\tabcolsep}{4.5pt}
\begin{tabular}{l cccccc}
\toprule
\textbf{Strategy}
& \textbf{@10} & \textbf{@20} & \textbf{@25} & \textbf{@30} & \textbf{@35} & \textbf{@40} \\
\midrule
Single (Qwen)
& 0.365 & 0.306 & 0.280 & 0.222 & 0.130 & 0.040 \\
Majority Vote
& 0.365 & 0.305 & 0.280 & 0.223 & 0.130 & 0.041 \\
Uniform
& 0.365 & 0.305 & 0.280 & 0.222 & 0.131 & 0.041 \\
Confidence
& 0.365 & 0.306 & 0.280 & 0.223 & 0.130 & 0.037 \\
TASRE ($T{=}1.0$)
& 0.365 & 0.305 & 0.281 & 0.221 & 0.129 & 0.040 \\
\midrule
\textit{Max $-$ Min}
& \textit{0.000} & \textit{0.001} & \textit{0.001} & \textit{0.002} & \textit{0.002} & \textit{0.004} \\
\bottomrule
\end{tabular}}
\end{table}

All five strategies produce nearly indistinguishable results.
The largest inter-strategy variation occurs at $\tau = 40$~dBZ (CSI range $= 0.004$),
which is within stochastic training noise. The mean CSI spread across all strategies
is only $0.0007$, confirming that the choice of fusion method has negligible impact
on final retrieval performance.

\paragraph{Experiment 2: TASRE Temperature Ablation.}

Table~\ref{tab:tasre_temp} reports the effect of the softmax temperature $T$ on
TASRE-fused training.
\begin{table}[h]
\centering
\caption{TASRE temperature ablation.
  Mean metrics averaged over all six thresholds.
  \textbf{Bold}: best per column.}
\label{tab:tasre_temp}
\resizebox{0.5\textwidth}{!}{
\vspace{4pt}
\small
\setlength{\tabcolsep}{4pt}
\begin{tabular}{l cccc}
\toprule
\textbf{Temperature}
& \textbf{CSI-Mean}$\uparrow$
& \textbf{FAR-Mean}$\downarrow$
& \textbf{POD-Mean}$\uparrow$
& \textbf{HSS-Mean}$\uparrow$ \\
\midrule
$T = 0.5$
& 0.2238 & 0.4709 & 0.2716 & 0.3255 \\
$T = 1.0$
& 0.2239 & 0.4701 & 0.2720 & 0.3259 \\
$T = 2.0$
& 0.2232 & 0.4717 & 0.2710 & 0.3245 \\
\midrule
\textit{Max $-$ Min}
& \textit{0.0007} & \textit{0.0016} & \textit{0.0010} & \textit{0.0014} \\
\bottomrule
\end{tabular}}
\end{table}

\begin{table}[h]
\centering
\caption{Per-threshold CSI for TASRE temperature ablation.
  \textbf{Bold}: best per column.}
\label{tab:tasre_temp_per_threshold}
\resizebox{0.5\textwidth}{!}{
\vspace{4pt}
\small
\setlength{\tabcolsep}{4.5pt}
\begin{tabular}{l cccccc}
\toprule
\textbf{Temperature}
& \textbf{@10} & \textbf{@20} & \textbf{@25} & \textbf{@30} & \textbf{@35} & \textbf{@40} \\
\midrule
$T = 0.5$
& 0.365 & 0.305 & 0.280 & 0.222 & 0.130 & 0.041 \\
$T = 1.0$
& 0.365 & 0.305 & 0.281 & 0.221 & 0.129 & 0.039 \\
$T = 2.0$
& 0.365 & 0.305 & 0.280 & 0.222 & 0.128 & 0.039 \\
\midrule
\textit{Max $-$ Min}
& \textit{0.000} & \textit{0.000} & \textit{0.001} & \textit{0.001} & \textit{0.002} & \textit{0.002} \\
\bottomrule
\end{tabular}}
\end{table}

Temperature variation has virtually no effect: the CSI-Mean spread is 0.0007
and the maximum per-threshold CSI difference is 0.002. This is expected given
that the three VLM sources have nearly identical CSI scores
(Table~\ref{tab:vlm_csi}), making the softmax temperature irrelevant.

\paragraph{Summary.}

Table~\ref{tab:vlm_csi} reveals the central finding: \textbf{all three VLMs
produce annotations of nearly identical downstream quality} (CSI standard
deviation $< 0.0004$). Consequently:

\begin{enumerate}[nosep,leftmargin=1.5em]
  \item The TASRE weights converge to approximately uniform ($\sim$0.333 each),
    making TASRE indistinguishable from the Uniform strategy.
  \item Temperature has negligible effect on the weight distribution, as
    $\exp(\mathrm{CSI}_m / T) \approx \exp(\mathrm{CSI}_{m'} / T)$ when the CSI
    scores are tightly clustered.
  \item All five fusion strategies produce functionally equivalent pseudo-labels,
    since the three annotation sources agree to within noise.
\end{enumerate}

These results carry important implications:

\paragraph{Justification for Single-VLM (V2) as the Main Experiment.}
Since multi-VLM fusion offers no measurable improvement over single-source Qwen
annotations, the added complexity of the TASRE pipeline---querying multiple VLM
APIs, computing fusion weights, maintaining multiple annotation files---constitutes
\textbf{over-engineering} for this task. The main experiment therefore uses V2
(Qwen-only), which achieves the same downstream performance with a simpler
pipeline.

\paragraph{Why VLMs Agree.}
The five-attribute taxonomy (cloud type, convective depth, storm mode,
overshooting top, cirrus shield) defines a structured, constrained labelling
space. Modern VLMs, trained on overlapping corpora with similar visual reasoning
capabilities, converge to similar predictions in this well-defined categorical
space. The inter-model variance is dominated by stochastic sampling rather than
systematic disagreement.

\paragraph{When Multi-VLM Fusion May Help (Future Work).}
We hypothesise that TASRE-style fusion would become valuable in settings where:
\begin{itemize}[nosep,leftmargin=1.5em]
  \item The attribute taxonomy is \textbf{more fine-grained}, e.g., distinguishing
    sub-types of MCS or multiple cirrus morphologies, where VLM biases diverge.
  \item The task requires \textbf{open-ended text generation} rather than
    categorical selection, amplifying inter-model variance.
  \item VLMs are applied to \textbf{novel domains} (e.g., planetary atmospheres)
    where pre-training coverage varies substantially across models.
\end{itemize}

We consider the design of more refined multi-VLM fusion strategies that can
exploit such inter-model diversity as a promising direction for future work.

\begin{figure*}[t]
    \centering
    \includegraphics[width=\textwidth]{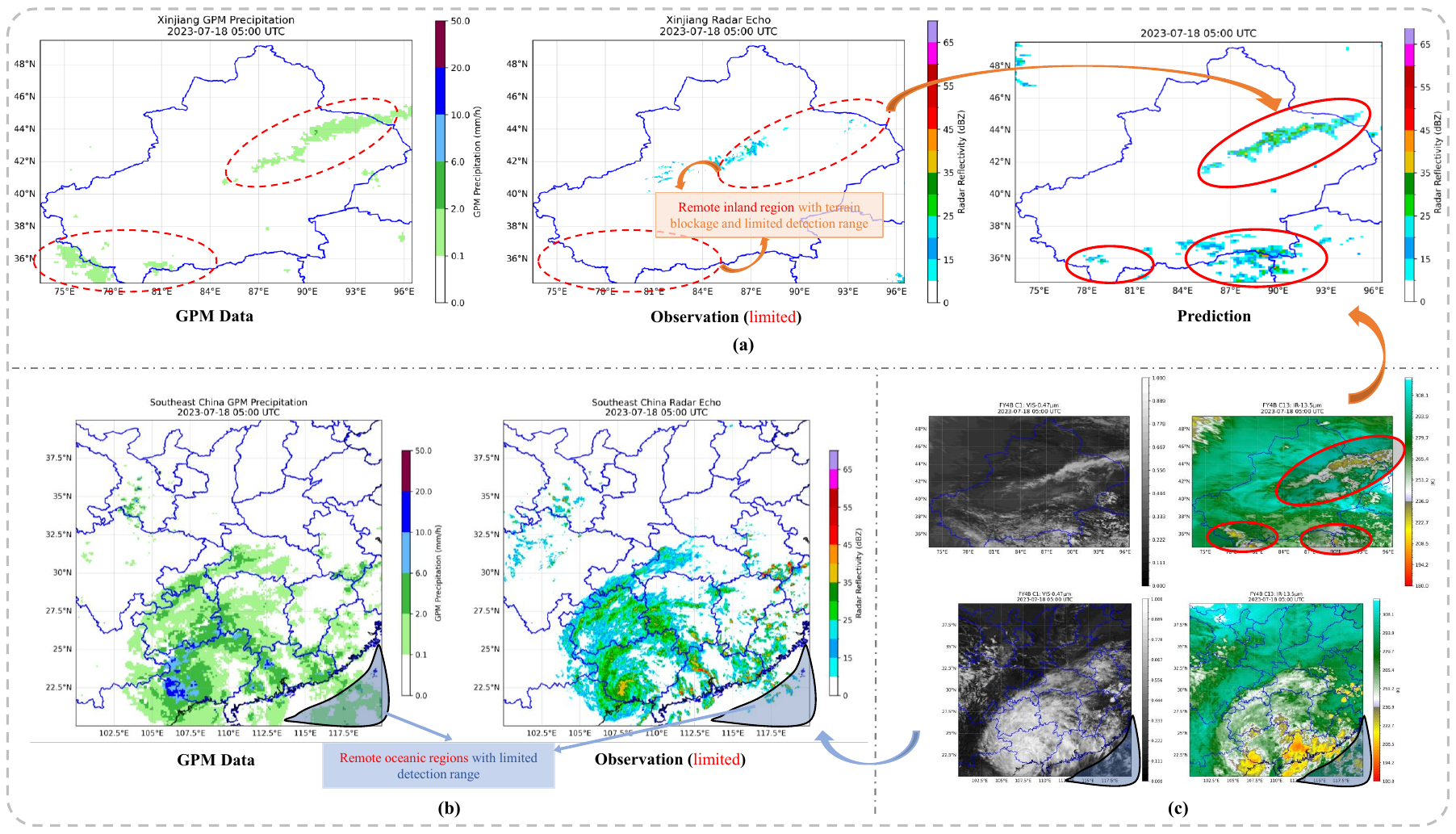}
    \caption{
        Practical applications of satellite-based radar retrieval in observationally-sparse regions. \textbf{(a)} Cross-regional transfer from southeastern China to Xinjiang Province, demonstrating retrieval capability in terrain-blocked mountainous regions through comparison of GPM precipitation data, limited ground radar observations, and our prediction results. \textbf{(b)} Oceanic coverage enhancement in southeastern China coastal areas where ground-based radar detection range is insufficient. \textbf{(c)} Multi-source satellite observations (VIS and IR channels) for both regions, showing convective systems visible in satellite imagery over southern Xinjiang where no ground radar coverage exists. Red circles highlight regions with limited radar coverage.
    }
    \label{fig:ca}
  \vspace{-15pt}
\end{figure*}

\subsection{Practical Significance of Satellite-Based Radar Retrieval}

Ground-based weather radar networks, while providing high-resolution precipitation observations, face significant operational limitations that restrict comprehensive meteorological monitoring. Our satellite-based radar retrieval framework addresses these critical observational gaps, offering a complementary solution for enhanced weather surveillance.

As demonstrated in Fig.~\ref{fig:ca}, ground-based radar networks face inherent limitations including terrain blockage in mountainous regions, limited detection range over oceanic areas, and sparse coverage in remote locations. These observational gaps pose significant challenges for severe weather monitoring and early warning systems.

Fig.~\ref{fig:ca}(a) illustrates the application of our method to Xinjiang Province, China, a high-altitude mountainous region where radar observations are severely limited due to topographic blockage and sparse radar station deployment. Notably, the southern regions of Xinjiang completely lack ground radar coverage, yet satellite observations in Fig.~\ref{fig:ca}(b) and Fig.~\ref{fig:ca}(c) clearly reveal active convective systems in these areas. The comparison between GPM precipitation data (10 km resolution satellite product), limited ground radar observations, and our retrieval results demonstrates successful domain adaptation capability. Our model effectively reconstructs precipitation patterns in these radar-blind regions, successfully retrieving convective activity that is visible in satellite imagery but undetectable by ground-based radar networks.


\subsection{Analysis and Discussion}

\noindent\textbf{Is a single VLM annotator sufficient for Phase~2a?}

Phase~2a initialises $\pi_\phi$ on pseudo-labels produced by a frozen
VLM, so annotation quality directly sets the starting point for GRPO.
A practical concern is whether relying on a single source
(Qwen-VL-Max) introduces systematic labelling bias that a
multi-source ensemble could correct.
We test this via a controlled study under the
\textbf{TASRE} (Task-Aligned Source Reliability Evaluation)
framework (Appendix~\ref{app:multi_vlm}), which weights each
annotation source by its downstream retrieval skill measured
through the frozen $U_\theta$:
\begin{equation}
  w_m = \frac{\exp(\mathrm{CSI}_m / \tau_w)}
             {\sum_{j} \exp(\mathrm{CSI}_j / \tau_w)},
  \label{eq:tasre_main}
\end{equation}
where $\mathrm{CSI}_m$ is the mean CSI obtained when training
$\pi_\phi$ on source~$m$ exclusively, and $\tau_w$ is a softmax
temperature controlling weight sharpness.
We evaluate three VLMs---Qwen-VL-Max, Kimi-VL, and
DeepSeek-VL---under five fusion strategies (Single, Majority Vote,
Uniform, Confidence-weighted, and TASRE), with results in
Table~\ref{tab:tasre_summary}.

\begin{table}[h]
\centering
\caption{Multi-VLM annotation fusion strategies evaluated on
  the FY-4B test set. Metrics are averaged over thresholds
  $\tau \in \{10,20,25,30,35,40\}$~dBZ under identical
  GRPO configuration (Appendix~\ref{app:multi_vlm}).}
\label{tab:tasre_summary}
\resizebox{0.48\textwidth}{!}{
\setlength{\tabcolsep}{5pt}
\renewcommand{\arraystretch}{1.15}
\begin{tabular}{lcccc}
\toprule
\textbf{Strategy} & \textbf{CSI-Mean}$\uparrow$ &
\textbf{FAR-Mean}$\downarrow$ & \textbf{POD-Mean}$\uparrow$ &
\textbf{HSS-Mean}$\uparrow$ \\
\midrule
Single (Qwen)          & 0.2238 & 0.4756 & 0.2719 & 0.3254 \\
Majority Vote          & 0.2241 & 0.4697 & 0.2722 & 0.3262 \\
Uniform                & 0.2240 & 0.4700 & 0.2721 & 0.3259 \\
Confidence             & 0.2234 & 0.4726 & 0.2713 & 0.3249 \\
TASRE ($\tau_w{=}1.0$) & 0.2239 & 0.4701 & 0.2720 & 0.3259 \\
\midrule
\textit{Max$-$Min}
  & \textit{0.0007} & \textit{0.0059} & \textit{0.0009}
  & \textit{0.0013} \\
\bottomrule
\end{tabular}}
\end{table}

All three VLMs produce annotations of essentially identical
downstream quality: the inter-source CSI standard deviation
is below~0.0004 (Appendix Table~XI), and the five fusion
strategies are indistinguishable within stochastic training
noise (CSI-Mean spread\,$=$\,0.0007).
The underlying reason is structural.
The five-attribute taxonomy imposes a tightly constrained
categorical label space---four cloud types, five storm modes,
and three binary-valued convective indicators---that leaves
little room for systematic inter-model disagreement among
modern VLMs sharing overlapping pre-training corpora; as
Appendix Table~XI confirms, all three sources cluster near
$\mathrm{CSI}\!=\!0.242\pm0.0004$, so TASRE weights
collapse to $\approx\!0.333$ regardless of $\tau_w$.

Two decisions follow directly.
\textit{First}, varying $\tau_w \in \{0.5, 1.0, 2.0\}$ has
negligible effect (CSI-Mean spread\,$=$\,0.0007 across
temperatures; Appendix Table~XV): when source CSI scores
are this tightly clustered, the exponential terms in
Eq.~\eqref{eq:tasre_main} are trivially equal, reducing
TASRE to the Uniform strategy regardless of $\tau_w$.
\textit{Second}, the overhead of querying multiple VLM APIs,
managing per-source annotation files, and estimating fusion
weights constitutes over-engineering that yields no measurable
retrieval benefit for the present task.
We therefore adopt a \textbf{single Qwen-VL-Max annotator}
throughout the main experiment.

This equivalence is specific to the current constrained
taxonomy.
In settings with finer-grained label spaces---such as
distinguishing MCS sub-types or multiple cirrus
morphologies---predictions across VLMs are more likely to
diverge, at which point source-reliability weighting may
provide a genuine signal.
The same applies to novel observation domains with uneven
model pre-training coverage, such as planetary or high-latitude
atmospheres.
We identify TASRE-style fusion under these conditions as an
open direction for future work.

\subsection{Disscusion}
The current approach learns coarse-grained semantic attributes without explicit spatial grounding; integrating position encoding and local region-specific context may improve precipitation localization and morphology fidelity. Moreover, the fixed-size patch-based design restricts scalability to higher-resolution or full-domain inference; efficient semantic injection mechanisms (e.g., neural operators) are needed to maintain computational tractability.

\end{document}